\documentclass[10pt,twocolumn]{IEEEtran}
\usepackage{times}
\usepackage[final]{graphicx}
\usepackage{amsmath,amsfonts}
\usepackage{amssymb,amsbsy}
\usepackage{times}

\usepackage{epsf}

\newtheorem{lem}{Lemma}
\newtheorem{rem}{Remark}
\newtheorem{defn}{Definition}
\newtheorem{prp}{Proposition}
\newtheorem{thm}{Theorem}
\newtheorem{cor}{Corollary}

\newcommand\ignore[1]{}

\newcommand{\hsp}{\hspace{0.05in} }

\newcommand{\hspp}{\hspace{0.02in} }
\newcommand{\hsppp}{\hspace{0.01in} }

\newcommand{\bH}{\mathbf{H}  }
\newcommand{\bQ}{\mathbf{Q}}
\newcommand{\bEe}{{\mathbf{E}}}

\newcommand{\lp}{{\ell^{'}}}
\newcommand{\mpr}{{m^{'}}}

\newcommand{\by}  { {\mathbf{y}}  }

\newcommand{\bx}  { {\mathbf{x}}  }

\newcommand{\ud} { {\mathrm{d}}  }  
\newcommand{\bw} { {\mathbf{w}} }
\newcommand{\bI} { {\mathbf{I}} }

\newcommand{\snr}{{\sf{SNR}} }

\newcommand{\mseq}{{\sf{MSE}} }

\newcommand{\ord}{{\mathcal{O}}}
\newcommand{\littleo}{{\mathnormal{o}}}

\newcommand{\sebnomin} {\mathnormal{  {\frac{E_b } {N_0}_{ {\mathrm{min}} }}  } }

\newcommand{\Tcoh}{\mathnormal{T_{coh}}}
\newcommand{\Wcoh}{\mathnormal{W_{coh}}}
\newcommand{\Ncoh}{\mathnormal{N_{c}}}

\newcommand{\nc}{\mathnormal{N_{c}}}

\newcommand{\eff}{{\mathrm{eff}} }

\newcommand{\sparse}{{\mathrm{sparse}} }
\newcommand{\rich}{{\mathrm{rich}} }
\newcommand{\peaky}{{\mathrm{peaky}} }

\newcommand{\fone}{\mathnormal{f_{1}}}
\newcommand{\ftwo}{\mathnormal{f_{2}}}
\newcommand{\fthr}{\mathnormal{f_{3}}}
\newcommand{\gone}{\mathnormal{g_{1}}}

\newcommand{\etr} { E_{tr} }
\newcommand{\Dwmax} { D_{W,\max} }
\newcommand{\Dtmax} { D_{T,\max} }

\newcommand{\etas} {\eta^{*}}

\newcommand{\tsty} {\textstyle}

\begin{document}
\renewcommand{\textfraction}{0}

\title{Capacity of Sparse Multipath Channels in the Ultra-Wideband Regime}
\author{ {Vasanthan Raghavan$^*$, Gautham Hariharan and Akbar M. Sayeed}
\thanks{Manuscript received XXXX; revised XXXX. This work was supported
in part by the National Science Foundation under Grant CCF-0431088.
This paper was presented in part at the $43$rd Allerton Conference
(Allerton 2005), Monticello, IL and at the $14$th Adaptive Sensor
Array Processing Workshop (ASAP 2006), Cambridge, MA. The associate
editor coordinating the review of this manuscript and approving it for
publication was Dr. Michael L. Honig.}
\thanks{V. Raghavan is with the Department of Electrical and
Computer Engineering, University of Illinois, Urbana-Champaign,
Urbana, IL 61801 USA (email: vasanthan\_raghavan@ieee.org).}
\thanks{G. Hariharan and A. M. Sayeed are with the Department of Electrical
and Computer Engineering, University of Wisconsin, Madison, WI
53706 USA (email: gauthamh@cae.wisc.edu, akbar@engr.wisc.edu).}
}

\date{}
\maketitle

\begin{abstract}
This paper studies the ergodic capacity of time- and
frequency-selective multipath fading channels in the ultrawideband
(UWB) regime when training signals are used for channel estimation
at the receiver. Motivated by recent measurement results on UWB
channels, we propose a model for sparse multipath channels. A key
implication of sparsity is that the independent degrees of freedom
(DoF) in the channel scale sub-linearly with the signal space
dimension (product of signaling duration and bandwidth). Sparsity
is captured by the number of resolvable paths in delay and
Doppler. Our analysis is based on a training and communication
scheme that employs signaling over orthogonal short-time Fourier
(STF) basis functions. STF signaling naturally relates sparsity in
delay-Doppler to coherence in time-frequency. We study the impact
of multipath sparsity on two fundamental metrics of spectral
efficiency in the wideband/low-SNR limit introduced by Verdu:
first- and second-order optimality conditions. Recent results by
Zheng {\em et. al.} have underscored the large gap in spectral
efficiency between coherent and non-coherent extremes and the
importance of channel learning in bridging the gap. Building on
these results, our results lead to the following implications of
multipath sparsity: 1) The coherence requirements are shared in
both time and frequency, thereby significantly relaxing the
required scaling in coherence time with SNR; 2) Sparse multipath
channels are asymptotically coherent --- for a given but large
bandwidth, the channel can be learned perfectly and the coherence
requirements for first- and second-order optimality met through
sufficiently large signaling duration; and 3) The requirement of
peaky signals in attaining capacity is eliminated or relaxed in
sparse environments.
\end{abstract}

\section{Introduction}
Emerging applications of ultrawideband (UWB) radio technology have
inspired both academic and industrial research on wide-ranging
problems. The large bandwidth of UWB systems results in
fundamentally new channel characteristics as evident from recent
measurement campaigns \cite{molisch,karedal,chiachin_residential}.
This is due to the fact that, analogous to radar, wideband
waveforms enable multipath resolution in delay at a much finer
scale -- delay resolution increases in direct proportion to
bandwidth. From a communication-theoretic perspective, the number
of resolvable multipath components reflects the number of
independent degrees of freedom (DoF) in the channel
\cite{akbar_and_venu,akbar_behnaam}, which in turn governs
fundamental limits on performance. When the channel coefficients
corresponding to resolvable multipath are perfectly known at the
receiver (coherent regime), the DoF reflect the level of
delay-Doppler diversity afforded by the channel
\cite{akbar_and_venu,ke_tamer_say}. On the other hand, when the
channel coefficients are unknown at the receiver (non-coherent
regime), the DoF reflect the level of uncertainty in the channel.
The fundamental limits to communication, such as capacity, can be
radically different in the coherent and non-coherent extremes, and
communication schemes that explicitly or implicitly learn the
channel can bridge the gap between the extremes
\cite{verdu,zheng}.

In this paper, we study the ergodic capacity of time- and
frequency-selective UWB channels in the non-coherent regime where
the channel is explicitly estimated at the receiver using training
signals. Motivated by recent measurement results, our focus is on
channels that exhibit {\em sparse} multipath -- the number of DoF
in the channel scale {\em sub-linearly} with the signal space
dimension (product of signaling duration and bandwidth) -- in
contrast to the widely prevalent assumption of rich multipath in
which the number of DoF scale linearly with signal space
dimension.  Whether a multipath channel is rich or sparse depends
on the operating frequency, bandwidth and the scattering
environment \cite{molisch}. For example, \cite{karedal} reports
rich channels even for $7.5$ GHz bandwidth in industrial
environments whereas \cite{chiachin_residential} reports sparse
multipath in residential environments at the same bandwidth.
Overall, large bandwidths increase the likelihood of channel
sparsity \cite{molisch,molisch_etal}. In time-selective scenarios,
the likelihood of sparsity is increased further due to multipath
resolution in Doppler.

The results in this paper build on two recent works that explore
ergodic capacity of fading channels in the wideband/low-SNR
regime~\cite{verdu,zheng}. The seminal work in~\cite{verdu} shows
that spectral efficiency in the wideband regime is captured by two
fundamental metrics: $\sebnomin$, the minimum energy per bit for
reliable communication, and $S_0$, the wideband slope. A signaling
scheme that achieves $\sebnomin$ is termed {\emph{first-order
optimal}} and one that achieves $S_0$ as well is termed
{\emph{second-order optimal}}. The results of \cite{verdu} also
show that knowledge of channel state information (CSI) at the
receiver imposes a sharp cut-off on the achievability of ergodic
capacity in the wideband regime.  In particular, while QPSK
signaling is second-order optimal when perfect CSI is available
(coherent regime), flash (peaky) signaling is necessary for
first-order optimality when no CSI is available (non-coherent
regime). However, a flash signaling scheme, besides having an
unbounded peak-to-average ratio (and hence practically
infeasible), also results in $S_0 = 0$ and thereby violating
second-order optimality.

This apparent sharp cut-off in the peak-to-average ratio of
capacity achieving signaling schemes between the coherent and
non-coherent extremes was examined in~\cite{zheng}. If the
coherence time of the channel scales at a sufficiently fast rate
with the bandwidth, Zheng {\em{et al.}} show that a communication
scheme with explicit training can bridge the gap between the two
extremes. However, no physical justification is provided for the
existence of such a scaling in coherence time with bandwidth. In
other related work, \cite{porrat} investigates the effect of
channel uncertainty when using spread-spectrum signals. They
conclude that the number of resolvable channel paths need to scale
sub-linearly with bandwidth in order to achieve the wideband limit
(first-order optimality in \cite{verdu}).

We first propose a model for sparse multipath channels to capture
the physical channel characteristics in the UWB regime as observed
in recent measurement studies. In a time- and frequency-selective
environment, multipath components can be resolved in delay and
Doppler where the resolution in delay/Doppler increases with
signaling bandwidth/duration \cite{akbar_behnaam}. A key
implication of multipath sparsity is that the number of DoF in the
channel (resolvable delay-Doppler channel coefficients) scales
sub-linearly with the signal space dimension. Our analysis of the
ergodic capacity of doubly-selective UWB channels is based on
signaling over short-time Fourier (STF) basis functions
\cite{kozek,ke_tamer_say} that are a generalization of OFDM
signaling and serve as approximate eigenfunctions for underspread
channels. Furthermore, STF signaling naturally relates multipath
sparsity in delay-Doppler to coherence or correlation in time and
frequency \cite{ke_tamer_say}. We consider a communication scheme
in which explicit training symbols are used to estimate the
channel at the receiver. The capacity of this scheme is then
studied to investigate the impact of multipath sparsity on
achieving coherent capacity.

The results of this paper lead to several new contributions and
insights on the impact of sparsity. First, we show that multipath
sparsity provides a natural physical mechanism for scaling of
coherence time, $T_{coh}$, with bandwidth/SNR, as assumed in
\cite{zheng}. Second, the coherence requirements for achieving
capacity are shared between both time and frequency: the coherence
bandwidth, $W_{coh}$, increases with bandwidth, $W$ (due to
sparsity in delay), and the coherence time, $T_{coh}$, increases
with signaling duration $T$ (due to sparsity in Doppler). As a
result, the scaling requirements on $T_{coh}$ with $W$ (or $\snr =
P/W$, where $P$ is the total transmit power) needed in
\cite{zheng} for first- and second-order optimality are replaced
by scaling requirements on the time-frequency coherence dimension
$N_{coh} = T_{coh} W_{coh}$. This leads to significantly relaxed
requirements on $T_{coh}$ scaling with bandwidth/SNR compared to
those in \cite{zheng}. Third, we show that sparse multipath
channels are {\em asymptotically coherent}; that is, for a
sufficiently large but fixed bandwidth, the conditions for first-
and second-order optimality can be achieved simply by making the
signaling duration sufficiently large. We quantify the required
(power-law) scaling in $T$ with $W$ for first- and second-order
optimality as a function of channel sparsity. This asymptotic
coherence of sparse channels is also manifested in the performance
of the training scheme -- consistent channel estimation is
achieved with vanishing fraction of energy expended on training.
The asymptotic coherence of sparse channels also
eliminates/relaxes the need for peaky signaling that has been
emphasized in existing results \cite{kennedy,verdu} on
non-coherent capacity, implicitly based on a rich multipath
assumption. We discuss how sparsity and peakiness can be traded
off suitably depending on system design requirements. Finally, the
results in this paper are shown to hold in general, independent of
the type of scaling laws used to model sparsity.

The paper is organized as follows. The system setup, including the
sparse channel model and training-based STF signaling scheme, is
described in Section~\ref{sec2}. In Section~\ref{sec3}, we study
the ergodic capacity of sparse channels with perfect CSI and for
the training-based communication scheme. A discussion of the
results, including their relation to existing work is provided in
Section~\ref{sec4}. Numerical results are provided to illustrate
the implications of the theoretical results. Concluding remarks
and directions for future work are discussed in
Section~\ref{sec5}.

\section{System Setup}
\label{sec2}

In this section, we first propose a model for sparse multipath
channels in terms of the number of paths that are resolvable in
delay and Doppler. We then develop a system model based on
orthogonal short-time Fourier (STF) signaling and propose a block
fading channel model that naturally relates multipath sparsity in
delay-Doppler to coherence in time-frequency. We then describe the
training-based communication scheme in the STF domain whose
capacity is investigated in this paper.

\subsection{Sparse Multipath Channel Modeling}
\label{sec2a} We consider a single-user single-antenna
communication system in complex baseband
\begin{eqnarray}
\label{io_eqn} y(t) = \int_{0}^{T_m} \! \! \!
\int_{-\frac{W_d}{2}}^{\frac{W_d}{2}}
 h(\tau,\nu) x(t-\tau) e^{j 2 \pi \nu t} \, \ud \nu \, \ud \tau
 + w (t)
\end{eqnarray}
where the channel is characterized by the delay-Doppler spreading
function, $h(\tau,\nu)$, and $x(t)$, $y(t)$ and $w(t)$ represent
the transmitted, received and additive white Gaussian noise (AWGN)
waveforms, respectively. $T_m$ and $W_d$ represent the delay and
Doppler spreads of the channel. We assume an underspread channel,
$T_m W_d \ll 1$, which is valid for most radio channels. A
physical discrete multipath channel can be modeled as
\begin{eqnarray}
h(\tau,\nu) & = & \sum_{n} \beta_n \delta(\tau-\tau_n) \delta(\nu
- \nu_n) \nonumber \\
y(t) & = & \sum_n \beta_n x(t-\tau_n)e^{j2\pi \nu_n t} + w(t)
\label{disc_mp}
\end{eqnarray}
where $\beta_n$, $\tau_n \in [0,T_m]$ and $\nu_n \in
[-W_d/2,W_d/2]$ denote the complex path gain, delay and Doppler
shift associated with the $n$-th path. Note that the above model
assumes that the carrier frequency is much larger than the
signaling bandwidth so that the effects of motion are accurately
captured via Doppler shifts (the shrinking or dilation of the
signaling waveforms is ignored).

\begin{figure*}[hbt!]
\begin{center}
\begin{tabular}{ccc}
\begin{minipage}{2.2in}
\centerline{\includegraphics[width=2.2in]{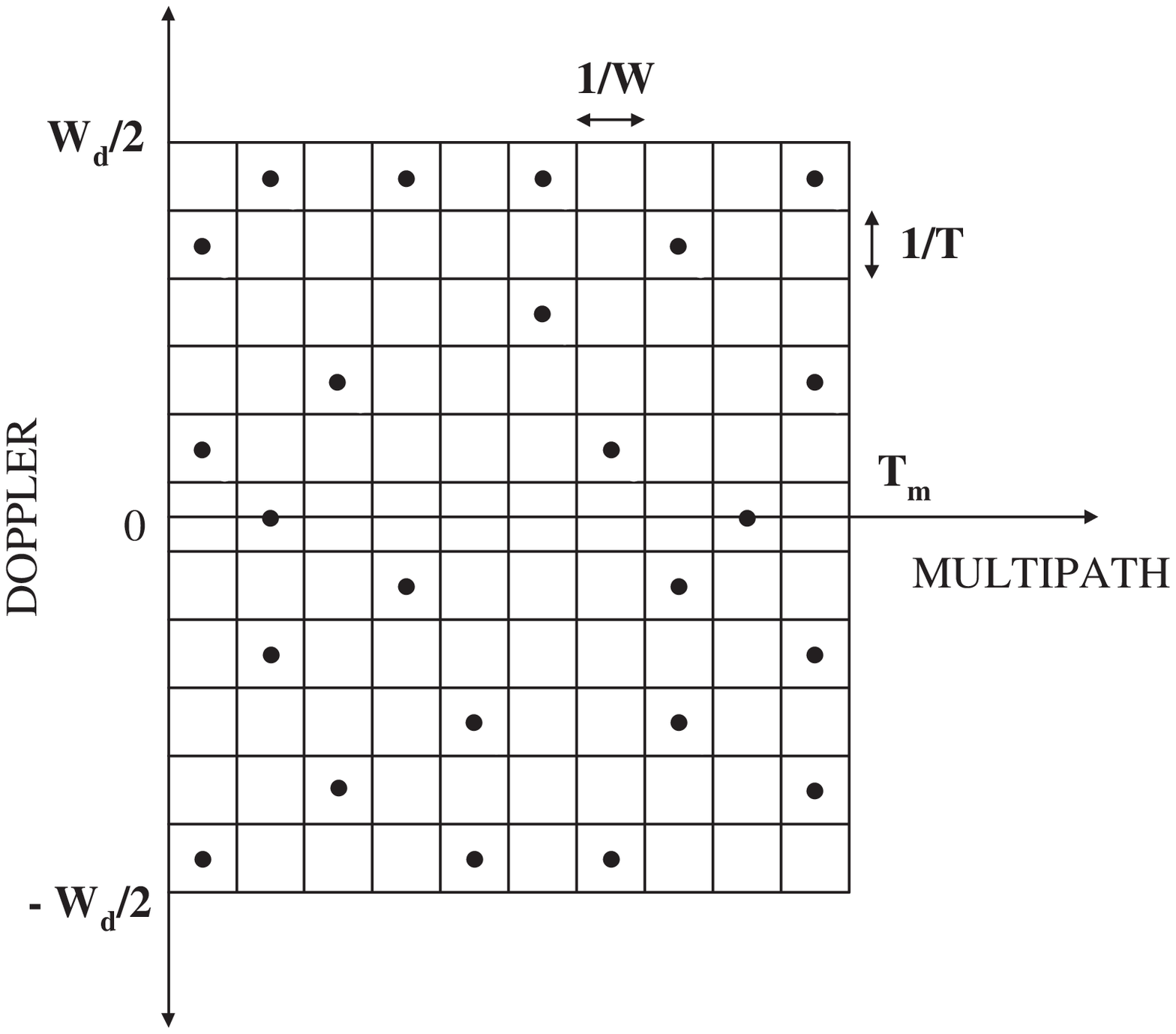}}
\end{minipage} &
\begin{minipage}{2.4in}
\centerline{\includegraphics[width=2.4in]{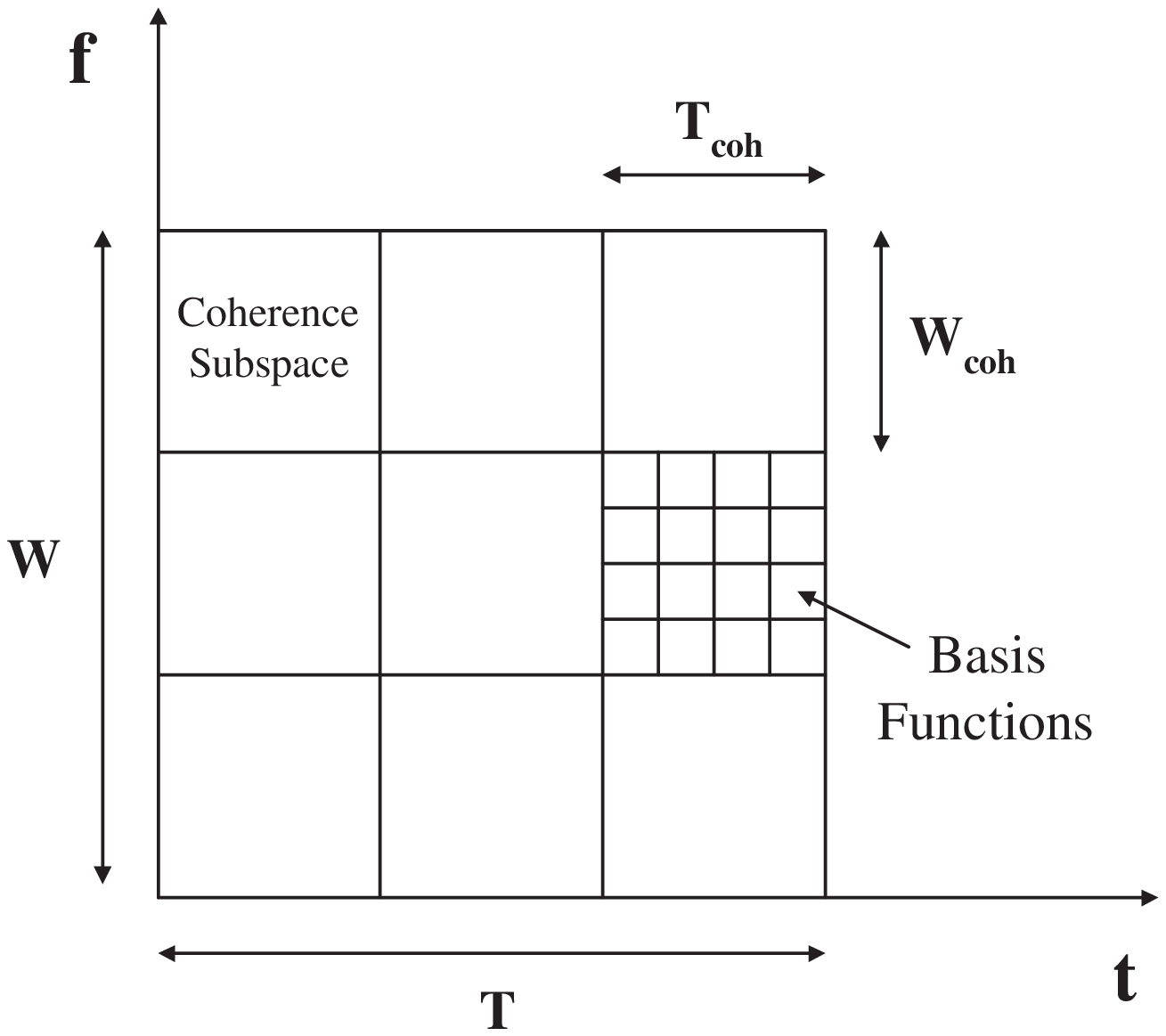}}
\end{minipage} &
\begin{minipage}{2.2in}
\centerline{\includegraphics[width=2.2in]{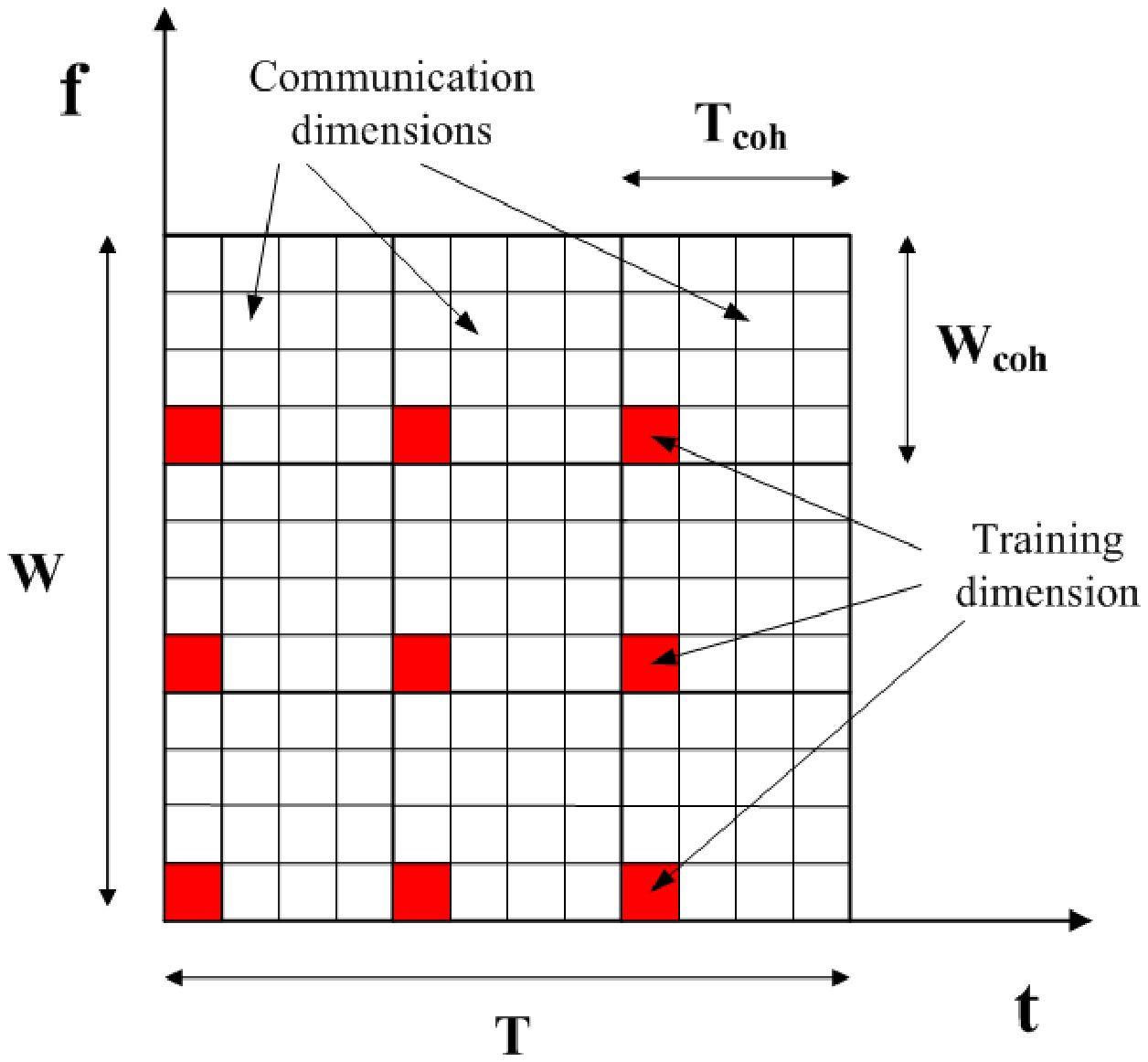}}
\end{minipage} \\
 (a) & (b) & (c)
\end{tabular}
\caption{ \label{fig:del_dopp} {\sl (a) Delay-doppler sampling
commensurate with signaling duration and bandwidth. (b)
Time-frequency coherence subspaces in STF signaling. (c)
Illustration of the training-based communication scheme in the STF
domain. One dimension in each coherence subspace (dark squares)
represents the training dimension and the remaining dimensions are
used for communication.}}
\end{center}
\vspace{-5mm}
\end{figure*}

The physical model (\ref{disc_mp}), while accurate, is complex to
analyze from a communication-theoretic perspective due to the
non-linear dependence on propagation parameters $\tau_n$ and
$\nu_n$. We instead use a linear \emph{virtual representation}
\cite{akbar_behnaam, akbar_and_venu} for time- and
frequency-selective multipath channels that captures the channel
characteristics in terms of {\em resolvable paths} and greatly
facilitates analysis. Throughout the paper, we consider signaling
over a duration $T$ and (two-sided) bandwidth $W$. The virtual
representation, illustrated in Fig.~\ref{fig:del_dopp}(a),
uniformly samples the multipath in delay and Doppler at a
resolution commensurate with $W$ and $T$, respectively
\cite{akbar_behnaam,akbar_and_venu}
\begin{eqnarray}
y(t) &=& \sum_{\ell=0}^{L} \sum_{m=-M}^{M}h_{\ell,m} x(t-
\ell/W)e^{j2\pi
mt/T}  \label{del_dopp_samp} \\
\ h_{\ell,m} &\approx&  \sum_{n \in S_{\tau,\ell} \cap S_{\nu,m}}
\beta_n  \label{hlm}
\end{eqnarray}
where $L = \lceil T_m W\rceil$, $M = \lceil TW_d/2\rceil$,
$S_{\tau,\ell} = \{ n: \ell/W - 1/2W< \tau_n \leq \ell/W + 1/2W\}$
denotes the set of all paths whose delays lie within the delay
resolution bin of width $\Delta \tau = 1/W$ centered around the
$\ell$-th resolvable (virtual) delay, $\hat{\tau} = \ell/W$, and
$S_{\nu,m} = \{n: m/T - 1/2T < \nu_n \leq m/T + 1/2T \}$ denotes
the set of all paths whose Doppler shifts lie within the Doppler
resolution bin of width $\Delta \nu = 1/T$ centered around the
$m$-th resolvable (virtual) Doppler shift, $\hat{\nu}_m = m/T$.
The sampled representation (\ref{del_dopp_samp}) is linear and is
characterized by the virtual delay-Doppler channel coefficients
$\{ h_{\ell,m} \}$. The expression (\ref{hlm}) states that the
channel coefficient $h_{\ell,m}$ consists of the sum of gains of
all paths whose delays and Doppler shifts lie within the
$(\ell,m)$-th delay-Doppler resolution bin of width $\Delta \tau
\times \Delta \nu$ centered around the sampling point
$(\hat{\tau},\hat{\nu}_m) = (\ell/W,m/T)$ in the $(\tau,\nu)$
(delay-Doppler) space. It follows that \emph{distinct}
$h_{\ell,m}$'s  correspond to approximately\footnote{Approximate due to
finite $T$ and $W$.} \emph{disjoint} subsets of paths and are hence
approximately statistically independent (due to independent path
phases). This approximation gets more accurate with increasing $T$
and $W$, due to higher delay-Doppler resolution, and we assume
that the channel coefficients $\{ h_{\ell,m}\}$ are perfectly
independent. We also assume Rayleigh fading in which $\{
h_{\ell,m}\}$ are zero-mean Gaussian random
variables.\footnote{This would be true if, for example, there are
sufficiently large number of {\em unresolvable} paths contributing
to each $h_{\ell,m}$ in (\ref{hlm}).} Thus, for Rayleigh fading,
the channel statistics are characterized by the power in the
virtual channel coefficients
\begin{equation}
\Psi(\ell,m) = E[|h_{\ell,m}|^2] \approx  \sum_{n \in
S_{\tau,\ell} \cap S_{\nu,m}} E[|\beta_n|^2]. \label{hlm_power}
\end{equation}

We define {\em dominant non-zero channel coefficients},
$h_{\ell,m}$'s, as those which contribute significant channel
power; that is, the coefficients for which $\Psi(\ell,m) > \gamma$
for some prescribed threshold $\gamma > 0$.\footnote{The choice of
the threshold $\gamma$ depends on the operating $\snr$ and
discussion of the choice of this threshold is beyond the scope of
this paper.} In Fig.~\ref{fig:del_dopp}(a), the delay-Doppler
resolution bins with a dot in them represent the dominant channel
coefficients. Let $D$ denote the number of dominant non-zero
channel coefficients; that is, $D = | \{ (\ell,m): \Psi(\ell,m) >
\gamma \}|$. The parameter $D$ reflects the (dominant)
statistically independent degrees of freedom (DoF) in the channel
and also signifies the delay-Doppler diversity afforded by the
channel. Furthermore, we decompose $D$ as $D  = D_T D_W$ where
$D_T$ denotes the Doppler/time diversity and $D_W$ the
frequency/delay diversity. The channel DoF or delay-Doppler
diversity is bounded as:
\begin{eqnarray}
D  =  D_{T} D_{W} \leq D_{\max} = \Dtmax \Dwmax \nonumber \\
\Dtmax = \left \lceil TW_d  \right \rceil \ , \ \Dwmax =
\left\lceil T_m W \right \rceil  \label{del_dopp_div}
\end{eqnarray}
where $\Dtmax$ denotes the maximum number of resolvable paths in
Doppler (maximum Doppler/time diversity) and $\Dwmax$ denotes
maximum number of resolvable paths in delay (maximum
delay/frequency diversity). Note that $\Dtmax$ and $\Dwmax$
increase linearly with $T$ and $W$, respectively. $D = D_{\max}$
represents a rich multipath environment in which each resolution
bin in Fig.~\ref{fig:del_dopp}(a) corresponds to a dominant
channel coefficient.

However, from recent measurement
campaigns~\cite{molisch,chiachin1,saadane} for UWB channels, there
is growing experimental evidence that the dominant channel
coefficients get sparser in delay as the bandwidth increases. Most
existing measurement results are for indoor UWB channels and do
not consider the effect of Doppler. We are interested in modeling
scenarios with Doppler effects, as well, due to motion. In such
cases, as we consider large bandwidths and/or long signaling
durations, the resolution of paths in both delay and Doppler
domains gets finer, leading to the scenario in
Fig.~\ref{fig:del_dopp}(a) where the delay-Doppler resolution bins
are sparsely populated with paths, i.e. $D < D_{\max}$.

We formally model multipath sparsity with a {\em sub-linear}
scaling in $D_T$ and $D_W$ with $T$ and $W$:
\begin{equation}
D_T \sim  (TW_d)^{\delta_1} \ , \ D_W \sim (T_mW)^{\delta_2} \ , \
\delta_1, \delta_2 \in [0,1] \label{sparse}
\end{equation}
where the smaller the value of $\delta_i$, the slower (sparser)
the growth in the resolvable paths in the corresponding domain.
Note that this directly implies that the total number of
delay-Doppler DoF, $D = D_T D_W$, scales sub-linearly with the
number of signal space dimensions $N=TW$.

\begin{rem}
\label{rem1} We focus on the power-law scaling in (\ref{sparse})
as a concrete example for studying the impact of sparsity on
capacity. As discussed in Sec.~\ref{sec4f}, the results of this
paper hold true for arbitrary sub-linear scaling laws.
\end{rem}
\begin{rem}
\label{rem2} With perfect CSI at the receiver, the parameter $D$
denotes the delay-Doppler diversity afforded by the channel,
whereas with no CSI, it reflects the level of channel uncertainty;
the number of channel parameters that need to be estimated for
coherent processing at the receiver.
\end{rem}

\subsection{Orthogonal Short-Time Fourier Signaling}
\label{sec2b} We consider signaling using an orthonormal
short-time Fourier (STF) basis~\cite{ke_tamer_say,kozek} that is a
natural generalization of orthogonal frequency-division
multiplexing (OFDM) for time-varying channels.\footnote{STF
signaling can be considered as OFDM signaling over a block of OFDM
symbol periods and with an appropriately chosen OFDM symbol
duration.} An orthogonal STF basis for the signal space is
generated from a fixed prototype waveform $g(t)$ via time and
frequency shifts: $\phi_{\ell m}(t) = g(t-\ell T_o)e^{j2\pi
W_ot}$, where $T_o W_o = 1$, $\;$ $\ell = 0,\cdots,N_{T}-1$, $\;$
$m=0,\cdots,N_{W}-1$ and $N = N_{T}N_{W} = TW$ with $N_{T} =
T/T_{o}, N_{W}=W/W_{o}$. The transmitted signal can be represented
as
\begin{equation}
x(t) = \sum \limits_{\ell=0}^{N_{T}-1} \sum
\limits_{m=0}^{N_{W}-1} x_{\ell m} \phi_{\ell m}(t) \hsp \hsp \,\
\hsp 0 \leq t \leq T \label{stf_mod}
\end{equation}
where $\{ x_{\ell m} \}$ represent the $N$ transmitted symbols
that are modulated onto the STF basis waveforms. At the receiver,
the received signal is projected onto the STF basis waveforms to
yield the received symbols
\begin{equation}
y_{\ell m} = \langle y, \phi_{\ell m} \rangle = \sum
\limits_{\ell^{'},m^{'}} h_{\ell m, \ell^{'} m^{'}} \;\
x_{\ell^{'} m^{'}} + w_{\ell m}. \label{stf_rx}
\end{equation}
We can represent the system using an $N$-dimensional matrix
equation
\begin{equation}
\by  =  \sqrt{\snr} \hsp {\mathbf{H}} \bx + \bw
\label{disc_channel}
\end{equation}
where $\bw$ represents the additive noise vector whose entries are
i.i.d.\ ${\mathcal{CN}}(0,1)$. The $N \times N$ matrix consists of
the channel coefficients $\{h_{\ell m, \ell^{'} m^{'}}\}$ in
(\ref{stf_rx}). The parameter $\snr$ represents the transmit
energy per modulated symbol and for a given transmit power $P$
equals $ \snr = \frac{TP}{TW} = \frac{P}{W}$. In this work, our
focus is on the UWB regime, where $\snr \rightarrow 0$ as $W
\rightarrow \infty$ for a fixed $P$.

For sufficiently underspread channels, the parameters $T_o$ and
$W_o$ can be matched to $T_m$ and $W_d$ so that the STF basis
waveforms serve as approximate eigenfunctions of the channel
\cite{kozek,ke_tamer_say}; that is, (\ref{stf_rx}) simplifies
to\footnote{The STF channel coefficients are different from the
delay-Doppler coefficients, even though we are using the same
symbols.} $y_{\ell m} \approx h_{\ell m} x_{\ell m} + w_{\ell m}$.
Thus the $N \times N$ channel matrix $\bH$ is approximately
diagonal. In this work, we assume that $\bH$ is exactly diagonal;
that is,
\begin{equation}
{\mathbf{H}}  =  {\mathrm{diag}} \Big[ \underbrace{ {h}_{1 1}
\cdots {h}_{1 \Ncoh}}_{  {\mathrm{Subspace}} \hsp 1}, \hsp
\underbrace { {h}_{2 1} \cdots {h}_{2 \Ncoh} }_{
{\mathrm{Subspace}} \hsp 2} \hsp  \cdots \hsp \underbrace { {h}_{D
1} \cdots {h}_{D \Ncoh} }_{ {\mathrm{Subspace}} \hsp D} \Big].
\label{H_diag}
\end{equation}

The diagonal entries of $\bH$ in (\ref{H_diag}) also admit an
intuitive block fading interpretation in terms of {\em
time-frequency coherence subspaces} \cite{ke_tamer_say}
illustrated in Fig.~\ref{fig:del_dopp}(b). The signal space is
partitioned as $N = TW=\nc D$ where $D$ represents the number of
statistically independent time-frequency coherence subspaces,
reflecting the DoF in the channel or the delay-Doppler diversity
(see (\ref{del_dopp_div})), and $\nc$ represents the dimension of
each coherence subspace, which we refer to as the
\textbf{coherence dimension}. In the block fading model in
(\ref{H_diag}), the channel coefficients over the $i$-th coherence
subspace $h_{i 1}, \cdots, h_{i \Ncoh}$ are assumed to be
identical, $\{ h_i \}$, whereas the coefficients across different
coherence subspaces are independent. Furthermore, due to the
stationarity of the channel statistics across time and frequency,
the different $h_i$ are identically distributed. Thus, the $D$
distinct STF channel coefficients, $\{ h_i \}$, corresponding to
the $D$ independent coherence subspaces, are i.i.d. zero-mean
Gaussian random variables (Rayleigh fading). The variance of each
channel coefficient is equal to $\bEe[|h_i|^2]= \sum_n
\bEe[|\beta_n|^2]$ which we normalize to unity
\cite{ke_tamer_say}.

Using the DoF scaling for sparse channels in (\ref{sparse}), the
coherence dimension can be computed as
\begin{eqnarray}
\Tcoh & = & \frac{T}{D_T} = T^{1-\delta_1}/W_d^{\delta_1}
\label{Tcoh} \\
\Wcoh & = & \frac{W}{D_W} = W^{1-\delta_2}/T_m^{\delta_2}
\label{Wcoh}\\
 \nc &= &\Tcoh \Wcoh
  = \frac{T^{1-\delta_1}}{W_d^{\delta_1}}
\frac{W^{1-\delta_2}}{T_m^{\delta_2}} \geq \left \lceil \frac{1}{
T_mW_d} \right \rceil \label{Ncoh}
\end{eqnarray}
where $\Tcoh$ is  the {\em coherence time} and $\Wcoh$ is the {\em
coherence bandwidth} of the channel, as illustrated in
Fig.~\ref{fig:del_dopp}(b). Note that $\delta_1 = \delta_2 = 1$
corresponds to a rich multipath channel in which $\nc = 1/(T_m
W_d)$ is \emph{constant} and $D = D_{\max}$ increases
\emph{linearly} with $N = TW$. This is the assumption prevalent in
existing works. In contrast, for sparse channels, $(\delta_1,
\delta_2) \in (0,1)$, and both $\nc$ and $D$ increase
{\emph{sub-linearly}} with $N$.

The coherence dimension plays a key role in our analysis. In terms
of channel parameters, $\nc$ increases with decreasing
$T_{m}W_{d}$ as well as with smaller $\delta_{i}$. In terms of
signaling parameters, $\nc$ can be increased by increasing $T$
and/or $W$. On the other hand, when the channel is rich, $\nc$
depends only on $T_{m}W_{d}$ and does not scale with $T$ or $W$.

Using (\ref{Wcoh}), we note that
\begin{equation}
\Wcoh = \frac{W^{1-\delta_2}}{(T_{m})^{\delta_2}} =
\frac{P^{1-\delta_2}}{(T_{m})^{\delta_2} \snr^{1-\delta_2}}
\label{Wcoh_snr}
\end{equation}
and thus $\Wcoh$ \emph{naturally} scales with $\snr$. Using
(\ref{Wcoh_snr}), the expression for $\nc$ in  (\ref{Ncoh}) becomes
\begin{equation}
\nc = \frac{T^{1-\delta_1}}{(W_{d})^{\delta_1}}
\frac{P^{1-\delta_2}}{(T_{m})^{\delta_2}\snr^{1-\delta_2}}.
\label{nc_wcoh}
\end{equation}
Our focus is on computing the sparse channel capacity and as we
will see later in Section \ref{sec3}, capacity turns out to be a
function only of the parameters $\nc$ and $\snr$. Furthermore, the
following relation between $\nc$ and $\snr = P/W$ plays a key role
in our analysis
\begin{equation}
\nc  =  \frac{1}{\snr^{\mu}} \hsp , \hsp \hsp \hsp \mu > 0 \hsp ,
\label{nc_snr_relation}
\end{equation}
where the parameter $\mu$ reflects the level of channel coherence.
Equating (\ref{nc_snr_relation}) with (\ref{nc_wcoh}) leads to the
following canonical relationship
\begin{equation}
T = \frac{\left( T_{m}^{\delta_2} W_{d}^{\delta_1}
\right)^{\frac{1}{1-\delta_1}} W^{\frac{\mu - 1 +
\delta_2}{1-\delta_1}}}{P^{\frac{\mu}{1-\delta_1}}}
\label{TWP_locus}
\end{equation}
that relates the signaling parameters ($T$,$W$,$P$), as a function
of the channel parameters, in order to satisfy
(\ref{nc_snr_relation}). Equations (\ref{nc_snr_relation}) and
(\ref{TWP_locus}) are the two key equations that capture the
impact of sparsity and we will revisit them in Section~\ref{sec4}.
We next describe the training-based communication scheme in the
STF domain that serves as the workhorse of the capacity analysis
in this paper.

\subsection{Training-Based Communication Using STF Signaling}
\label{sec2c} Our interest is primarily in the non-coherent
scenario when there is no CSI at the receiver \emph{a priori}. We
focus on a communication scheme in which the transmitted signals
include training symbols to enable channel estimation and coherent
detection. Although it is argued in
\cite{zheng},~\cite{hassibi_training} that training-based schemes
are sub-optimal from a capacity point of view, the restriction to
training schemes is motivated by practical considerations. We
assume that both the transmitter and the receiver have knowledge
of channel statistics (values of $T_m$, $W_d$, $\delta_1$ and
$\delta_2$ in our model).

We now describe the training-based communication scheme, adapted
from \cite{zheng} to STF signaling. The total energy available for
training and communication is $PT$, of which a fraction $\eta$ is
used for training and the remaining fraction $(1-\eta)$ is used
for communication. Since the quality of the channel estimate over
one coherence subspace depends only on the training energy and
{\emph{not}} on the number of training
symbols~\cite{hassibi_training}, our scheme uses one signal space
dimension in each coherence subspace for training and the
remaining $\left( \nc-1 \right)$ for communication, as illustrated
in Fig.~\ref{fig:del_dopp}(c). We consider minimum mean squared
error (MMSE) channel estimation and the two metrics that capture
channel estimation performance are (i) $\eta$, the fraction of
energy used for estimation, and (ii) $\mseq$, the mean squared
error in estimating each channel coefficient.

The training energy to estimate the channel coefficient in one
coherence subspace is given by
\begin{equation}
\etr = \frac{\eta TP}{D} \stackrel{\mathit{(a)}}{=} \eta \nc \snr
\label{E_tr}
\end{equation}
where (a) follows from the fact that $\snr = \frac{P}{W}$ and $\nc
D = TW$. Recall that $N = \nc D = TW = N_{T}N_{W}$ and
$D=D_{T}D_{W}$. Similarly we partition $\nc = N_{c,T}N_{c,W}$
where $N_{c,T} = N_T/D_T$ is the temporal coherence dimension and
$N_{c,W} = N_W/D_W$ is the spectral coherence dimension and
represent the number of STF basis functions that lie within
$T_{coh}$ and $W_{coh}$, respectively (see
Fig.~\ref{fig:del_dopp}(c)). The following equations describe
training in the STF system:
\begin{eqnarray}
y_{\ell m} & = & \sqrt{\etr } \hsp h_{\ell m} x_{\ell m}  + w_{\ell m}, \nonumber \\
  && \ell = (i-1)N_{c,T} + 1 \ , \  m = (j-1) N_{c,W} + 1,
 \nonumber \\
&&   i = 1, \cdots, D_T \ , \ j = 1, \cdots, D_W \label{training}
\end{eqnarray}
where $\{ x_{\ell m} \} $ are the $D$ training symbols (with  $|
x_{\ell m}|^2 = 1$) known at the receiver that are used to
estimate the $D$ channel coefficients $\{ h_{\ell m} \}$ with
${\bEe} [ | h_{\ell m} |^2 ] = 1$.

The communication energy per transmitted data symbol is given by
$E_{cm} = \frac{(1-\eta) T P }{(\nc-1)D} = \frac{(1-\eta) \nc \snr
}{(\nc-1)}$. The communication component of the system can be
described by
\begin{eqnarray}
y_{\lp \mpr} & = & \sqrt{E_{cm}} \hsp h_{\lp \mpr} x_{\lp \mpr}  + w_{\lp \mpr}, \nonumber \\
 \lp & = & (i-1)N_{c,T} + 2, \hsp \cdots, \hsp iN_{c,T} \nonumber \\  \mpr & = & (j-1) N_{c,W} +
  2, \hsp \cdots, \hsp jN_{c,W},
 \nonumber \\
&&   i = 1, \cdots, D_T \ , \ j = 1, \cdots, D_W \label{comm}
\end{eqnarray}
where $\{ x_{\lp \mpr} \}$ now represent the $(\nc-1) D$
communication symbols with ${\bEe} [ | x_{\lp \mpr} |^2 ]= 1$. We
can rewrite (\ref{comm}) as
\begin{align}
y_{\lp \mpr} = \sqrt{E_{cm}} \hsp \widehat{h}_{\lp \mpr} x_{\lp
\mpr} + \sqrt{E_{cm}} \hsp \Delta_{\lp \mpr} x_{\lp \mpr} + w_{\lp
\mpr} \label{comm_split}
\end{align}
and $ \widehat{h}_{\lp \mpr} $ is the MMSE estimate of $ h_{\lp
\mpr} $ and is given by
\begin{equation*}
\widehat{h}_{\lp \mpr} = \frac{\sqrt{\etr}}
 {1 + \etr  } \hsp  y_{\lp \mpr} x_{\lp \mpr}^*
\end{equation*}
and $\Delta_{\lp \mpr} = h_{\lp \mpr} - \widehat{h}_{\lp \mpr}$ is
the error in the estimate. The resulting $\mseq$ is given by
\begin{eqnarray}
\mseq(\eta,\nc,\snr) & = & {\bEe} \left[|h_{\lp \mpr} -
\widehat{h}_{\lp \mpr}|^{2} \right] \nonumber \\
& = & {\bEe} \left[ |\Delta_{\lp \mpr}|^{2} \right] \nonumber \\
& = & \frac{1}{1 + \etr } = \frac{1}{1+ \eta \nc \snr}. \label{MSE}
\end{eqnarray}
We are now ready to compute the ergodic capacity of the
training-based communication system.

\section{Ergodic Capacity of the Training-Based Communication Scheme}
\label{sec3} We first characterize the coherent capacity of the
wideband channel with perfect CSI at the receiver which serves as
a benchmark. The coherent capacity per dimension (in bps/Hz) is
\begin{equation}
C_{coh} \left(\snr \right) = \sup \limits_{ {\bQ}: \hsp
{\mathrm{Tr}}({\bQ}) \hsp \leq \hsp TP  } \frac{ {\bEe} \left[
\log_2 \det \left( \bI_{\nc D} + {\bH} {\bQ} {\bH}^H \right)
\right] }{\nc D}
\end{equation}
where $P$ denotes transmit power and $\bH$ is the diagonal,
block-fading channel matrix in (\ref{H_diag}). The optimization is
over the set of positive semi-definite transmit covariance
matrices $\bQ$. Due to the diagonal nature of $\bH$, the optimal
$\bQ$ is also diagonal. In particular, the uniform power
allocation $\bQ = \frac{TP} {\nc D} \bI_{\nc D} = \snr \hsp
\bI_{\nc D}$ achieves capacity and
\begin{eqnarray}
\label{exact_cap} C_{coh} \left( \snr \right) & = & \frac{
\sum_{i=1}^D {\bEe}\left[ \log_2 \left( 1 + \frac{TP}{\nc D}
\left| h_i \right|^2 \right) \right] }{D} \nonumber \\
& \stackrel{\mathit{(a)}}{=} &  {\bEe}\left[ \log_2 \left( 1 +
{\snr} \left| h \right| ^2 \right)  \right]
\end{eqnarray}
where (a) follows since $\{ h_i \}$ are i.i.d.\ with $h$
representing a generic random variable, $\nc D = TW$ and $\snr =
\frac{P}{W}$.

The next proposition provides upper and lower bounds to the
coherent capacity in the low $\snr$ regime.
\begin{prp}
\label{lowerbound} For all $b \in (0,1)$ and $\snr = \frac{P}{W}$
such that $\snr < \frac{(1-b)}{b}$, the coherent capacity
satisfies
\begin{eqnarray}
C_{coh} \left(\snr \right) & \geq & \log_2(e) \left(\snr - \snr^2
\right) \nonumber \\
C_{coh} \left( \snr \right) & \leq & \log_2(e) \left( \snr -
\frac{b}{2} \cdot \snr^2 \right). \label{coh_bounds}
\end{eqnarray}
Moreover the capacity converges to the lowerbound as $\snr
\rightarrow 0$.
\end{prp}
\begin{proof}
See Appendix \ref{append:app1}.
\end{proof}

The lowerbound in Proposition \ref{lowerbound} shows that the
minimum energy per bit for reliable communication is given by
$\sebnomin = \log_e(2)$ and the wideband slope $S_0 = 1$, the two
fundamental metrics defined in \cite{verdu}.

We now define the notion of an {\emph{operational coherence
level}} \cite{zheng} that allows an alternative, but equivalent,
characterization of capacity in the wideband/low-$\snr$ regime.
\begin{defn}
\label{defn_opcoh} Let $I_{tr}$ be the average mutual information
achievable with a training-based communication scheme. We say that
the scheme achieves an operational coherence level of $\epsilon$
$\left( 0 \leq \epsilon \leq 1 \right)$ if the low $\snr$ asymptote
of $I_{tr}$ is of the form $\snr - \ord \left( \snr^{1+\epsilon}
\right)$. Note that the two values of $\epsilon = 0$ and $\epsilon =
1$ correspond to the first-order and second-order optimality
conditions, respectively, as defined in \cite{verdu}.
\endproof
\end{defn}

In the scaling law, $\nc = \frac{1}{\snr^{\mu}}, \mu > 0$
in~(\ref{nc_snr_relation}), the parameter $\mu$ reflects the
coherence achieved by the training-based communication scheme. We
are interested in computing the value of $\mu$ such that the
training-based scheme achieves an operational coherence level of
$\epsilon$. This relation is characterized in Theorem~1. We start
with the following lemma that provides a lower bound to the
capacity of the training-based scheme.
\begin{lem}
\label{lem1}
The capacity of the training-based communication scheme described in
Sec.~\ref{sec2c} is lower bounded by
\begin{eqnarray}
I_{tr}(\eta, \nc, \snr) \geq \widehat{I}_{tr}(\eta,\nc,\snr)
\triangleq \frac{1}{2} \log_2 \left( 1 + 2\beta \sigma^2 \right)
\label{best_lower_bound}
\end{eqnarray}
where
\begin{equation}
\beta(\eta,\nc,\snr) =  \tsty{ \frac{ (1-\eta) \hspp (1 + \eta
\hspp \nc \snr) \hspp \nc \snr } {\left[ (\nc-1)(1 + \eta \hspp
\nc \snr) + (1-\eta) \hspp \nc \snr \right] } } \label{beta}
\end{equation}
\begin{equation}
\sigma^2(\eta,\nc,\snr) = \tsty{ \frac {\eta \hspp \nc \snr}{1 +
\eta \hspp \nc \snr } }. \label{sigmasq}
\end{equation}
\end{lem}
\vspace{0.07in}
\begin{proof}
See Appendix \ref{append:app2}.
\end{proof}

Next, we optimize over the fraction of energy spent for training,
$\eta$, to maximize the lower bound $\widehat{I}_{tr}$. Thus, we
explicitly highlight the role of $\eta$ in the following lemma.
\begin{lem}
\label{lem2} The $\eta$ that maximizes $\widehat{I}_{tr}\left(\eta
\right)$ given in (\ref{best_lower_bound}) satisfies $\frac{\ud
K(\eta) }{\ud \eta} = 0$ where $K(\eta) = K = \beta \sigma^2$ and
$\beta$ and $\sigma^2$  are as in (\ref{beta}) and
(\ref{sigmasq}), respectively. The optimizing value $\eta^{*}$ and
the corresponding $K^{*}$ are given by
\begin{eqnarray}
\etas & = & \tsty{ \frac{\nc \snr + \nc - 1}{(\nc-2) \nc
\snr}\cdot \left[ \sqrt{1 + \frac{\nc \snr (\nc-2)}{ \nc \snr +
\nc - 1}} - 1
\right] } \label{etaopt} \\
K^{*} & = & \tsty{ \frac{\nc \snr + \nc - 1}{(\nc-2)^{2}} \cdot
\left[ \sqrt{1 + \frac{\nc \snr (\nc-2)}{ \nc \snr + \nc - 1}} - 1
\right]^{2}}. \label{Kopt}
\end{eqnarray}
Furthermore, the optimized (tightest) lower bound is given by
\begin{equation}
\widehat{I}_{tr} \left( \etas \right) = \left(1-\frac{1}{\nc}
\right)\cdot \frac{1}{2} \cdot
\log_2 \left( 1 + 2 K^{*} \right). 
\label{I2opt}
\end{equation}
\end{lem}
{\vspace{0.07in}}
\begin{proof}
See Appendix \ref{append:app3}.
\end{proof}

We now state the main result of this work. The following theorem
characterizes the required scaling of $\nc$ (value of $\mu$) so
that any operational coherence level $\epsilon$ can be achieved.
\begin{thm}
{\label{thm1}} The average mutual information of the
training-based scheme achieves an operational coherence level
$\epsilon \in [0,1]$
\begin{eqnarray}
I_{tr} \geq \log_2(e) \cdot
\left[ \snr - \ord \left( \snr^{1+\epsilon} \right) \right]
\label{cap_tr}
\end{eqnarray}
if and only if $\nc = \frac{1}{\snr^{\mu}}$ for $\mu > 1 +
2\epsilon$. More precisely, if $\epsilon \in [0,1)$ and $N_c =
\frac{1}{\snr^{\mu}}$, $\mu > 1+2\epsilon = 1$, then
\begin{eqnarray}
I_{tr} \geq \log_2(e) \cdot \left[ \snr - 2 \hsp \snr^{1 +
\epsilon} + \littleo(\snr^{1 + \epsilon}) \right].
\label{epslessthanone}
\end{eqnarray}
If $\epsilon = 1$ and $N_c = \frac{1}{\snr^3}$, then
\begin{eqnarray}
I_{tr} \geq \log_2(e) \cdot \left[ \snr - 3 \hsp \snr^2 +
\littleo(\snr^2) \right]. \label{epsequalone}
\end{eqnarray}
If $\epsilon = 1$ and $N_c = \frac{1}{\snr^\mu}$, $\mu >
1+2\epsilon =3$, then
\begin{eqnarray}
I_{tr} \geq \log_2(e) \cdot \left[ \snr - \snr^2  +
\littleo(\snr^2) \right]. \label{mugreatthree}
\end{eqnarray}
In particular, the first- and second-order optimality conditions
(corresponding to $\epsilon = 0$ and $\epsilon = 1$) are met if and
only if $\mu > 1$ and $\mu > 3$, respectively.
\end{thm}
\begin{proof}
See Appendix \ref{append:app4}.
\end{proof}
Theorem~\ref{thm1} and equation (\ref{TWP_locus}) are key to
understanding the impact of sparsity on achieving coherent
capacity in the UWB regime. This is discussed in the next section.

\section{Discussion of Results}
\label{sec4}
\subsection{The Coherence Dimension: Sharing Coherence Costs in Time
and Frequency} \label{sec4a} 


Multipath sparsity provides a natural mechanism for channel
coherence and our results underscore the impact of sparsity in
both delay and Doppler via the notion of the time-frequency
coherence dimension, $\nc$. As discussed in Section \ref{sec2a},
in sparse channels, $D_W$ and $\Wcoh$ increase sub-linearly with
$W$. Furthermore, unlike existing works, we explicitly account for
Doppler diversity -- $D_T$ and $\Tcoh$ increase sub-linearly with
$T$ -- since STF signaling involves coding over multiple coherence
times.

\begin{figure}[hbt!]
\begin{center}
\begin{tabular}{c}
\begin{minipage}{3.45in}
\includegraphics[width=3.4in]{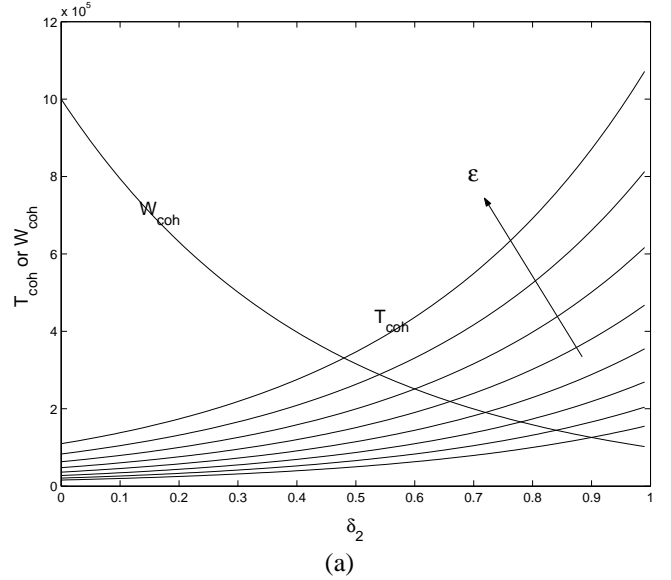} 
\end{minipage} \\ (a) \\
\begin{minipage}{3.45in}
\includegraphics[width=3.4in]{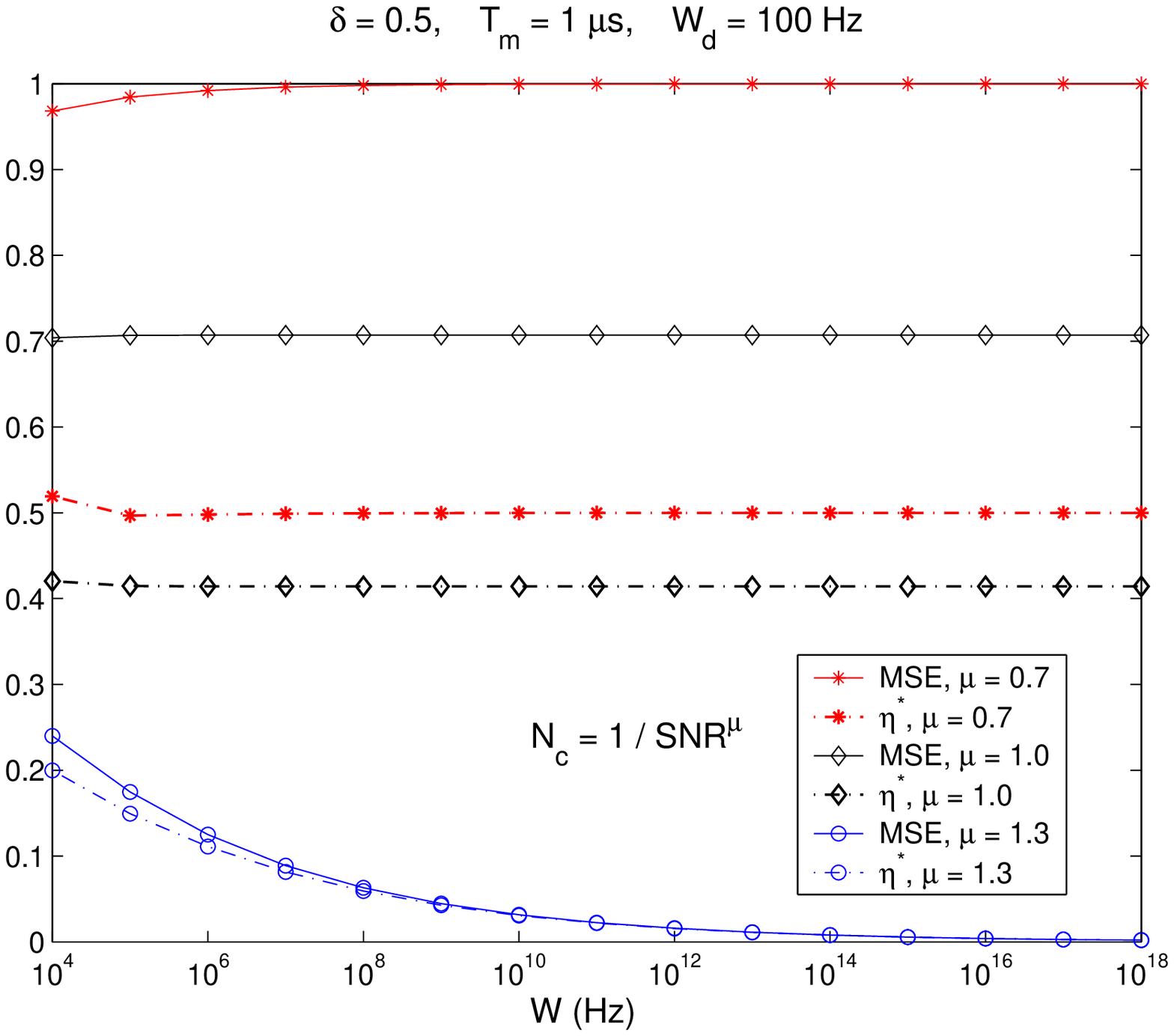}
\end{minipage} \\ (b) \\
\end{tabular}
\caption{\sl (a) The variation of $\Tcoh$ and $\Wcoh$ as a
function of delay sparsity ($\delta_2$). (b) $\mseq$ and $\etas$
for the channel estimation scheme as a function of $W$ for three
different values of $\mu$. } \label{fig_tcoh_mse}
\end{center}
\end{figure}

Theorem \ref{thm1} shows that the requirement on $\Tcoh$ in
\cite{zheng} is now the requirement on time-frequency coherence
dimension $\nc = \Tcoh \Wcoh$. Thus, the coherence cost is shared
in both time and frequency and as a result the required scaling
for $\Tcoh$ can be \emph{significantly weakened} by taking
advantage of the natural scaling of $\Wcoh$ with $W$. If the delay
diversity is known to scale as $D_W = \ord\left ( W^{\delta_2}
\right) \leftrightarrow \Wcoh = \ord \left( W^{1-\delta_2}
\right)$, then the $\Tcoh$ scaling requirement reduces to
\begin{equation}
\Tcoh = \nc/\Wcoh = \ord \left( W^{2 \epsilon +\delta_2} \right)
\label{tcohreq}
\end{equation}
to achieve an operational coherence level $\epsilon$, as per
Definition \ref{defn_opcoh}. For example, using $\epsilon = 0.5$,
which corresponds to a sub-linear term of $\snr^{1.5}$ in
(\ref{cap_tr}), and $\delta_2 = 0.5$, we get $\Tcoh = \ord(
W^{1.5} )$. This is a less stringent scaling law than would be
required using the framework of \cite{zheng}, where the
requirement would be $\Tcoh = \ord \left( W^{1+2 \epsilon} \right)
= \ord \left( W^{2} \right)$. The weaker $T_{coh}$ requirement for
sparse channels is graphically illustrated in
Fig.~\ref{fig_tcoh_mse}(a) for the following parameters: $T_{m} =
10^{-5}\hsp {\mathrm{secs.}}$, $W_{d} = 50 \hsp {\mathrm{Hz}}$, $W
= 50 \hsp {\mathrm{MHz}}$. Note that as the channel becomes more
sparse in delay (decreasing $\delta_2$), $W_{coh}$ gets larger,
thereby reducing the $T_{coh}$ requirement to achieve any desired
operational coherence $\epsilon$.

\subsection{Asymptotic Coherence of Sparse Channels}
\label{sec4b} Since channel uncertainty is the main factor that
affects capacity in the non-coherent scenario, we further
investigate the performance of channel estimation using two
metrics: (i) $\mseq$ of channel estimates and (ii) optimal
fraction of total energy used for estimation, $\etas$. The
following theorem characterizes the value of $\mu$ for
asymptotically energy-efficient and consistent estimation.
\begin{thm}
\label{thm2} In the limit of large signal space dimension ($T, W
\rightarrow \infty$)
\begin{equation}
\etas \rightarrow 0 \hsp \hsp \hsp \hsp \text{and} \hsp \hsp \hsp
\hsp \mseq = \frac{1}{1+\etr} \rightarrow 0 \label{asymp_eff}
\end{equation}
if and only if $\nc = \frac{1}{\snr^{\mu}} \hsp \hsp \hsp \hsp
\text{and} \hsp \hsp \hsp \hsp \mu > 1$.

Furthermore, the rates of convergence are given by
\begin{equation*}
\etas \rightarrow 0 \hsp \hsp \text{as} \hsp \hsp \ord \left(
\frac{1}{\sqrt{\nc \snr}} \right) = \ord \left( \snr^{\frac{\mu -
1}{2}} \right) = \ord \left( W^{\frac{1 - \mu}{2}} \right)
\end{equation*}
\begin{equation}
\etr \rightarrow \infty \hsp \hsp \text{as} \hsp \hsp \ord \left(
\sqrt{\nc \snr} \right) = \ord \left( \snr^{\frac{1 - \mu}{2}}
\right) = \ord \left( W^{\frac{\mu - 1}{2}} \right). \label{rates}
\end{equation}
\end{thm}
\vspace{0.1in}
\begin{proof}
See Appendix \ref{append:app5}.
\end{proof}
The above result says that multipath wireless channels are
\emph{asymptotically coherent} if and only if they are sparse and
$\nc$ satisfies the condition ($\mu > 1$) specified in Theorem
\ref{thm2}. For rich multipath, $\nc$ is a constant ($\nc =
\frac{1}{T_{m}W_{d}}$) and does not scale with $\snr$. For a
sparse channel with $\mu \leq 1$, $\nc$ does not scale at a fast
enough rate with $\snr$. Under both scenarios, as shown in the
proof of the theorem, the training scheme asymptotically uses half
the total energy ($\etas \rightarrow 0.5$) to estimate the channel
coefficients and the $\mseq$ does not decay to zero. For $\mu=1$,
the estimation performance is better than when $\mu < 1$, but
still not good enough to obtain asymptotic coherence. These
observations are illustrated in Fig.~\ref{fig_tcoh_mse}(b) where
$\etas$ and $\mseq$ are plotted as a function of increasing
bandwidth for three different cases: $\mu = 0.7$, $\mu=1$ and
$\mu=1.3$. In all the three cases, the signaling duration $T$ is
chosen according to (\ref{TWP_locus}).

Note that the requirement ($\mu > 1$) for asymptotic coherence in
Theorem~\ref{thm2} is exactly the same as the condition to achieve
first-order optimality in Theorem~\ref{thm1}. This makes intuitive
sense: with diminishing channel uncertainty ($\mseq \rightarrow
0$) and a vanishing fraction of the energy ($\etas \rightarrow 0$)
used for estimation, the capacity of the training-based system
converges to coherent capacity in the wideband limit.

\subsection{Optimal Choice of Signaling Parameters}
\label{sec4c} Recall the discussion in Section~\ref{sec2b}, in
particular equation (\ref{TWP_locus}) that relates the signaling
parameters ($T$,$W$,$P$) for achieving a desired scaling  of $\nc$
with $\snr$ in (\ref{nc_snr_relation}). We now revisit this
relationship, in light of Theorem~\ref{thm1}, and investigate the
choice of signaling parameters in order to obtain a desired level
of operational coherence $\epsilon$ (in particular, the values for
first- and second-order optimality, $\epsilon=0$ and $\epsilon=1$,
respectively).

Theorem~\ref{thm1} states that to achieve an operational coherence
$\epsilon$, the coherence dimension must scale as
\begin{equation}
\nc = \frac{1}{\snr^\mu} \ \ , \ \ \mu > 1 + 2\epsilon
\label{mu_eps}
\end{equation}
and by taking the logarithm of (\ref{TWP_locus}) we note that the signaling duration $T$
must scale with $W$ as a function of $P$ and the channel sparsity
parameters as
\begin{align}
& \log\left( T \right) = \frac{1}{1-\delta_1} \log \left(
W_{d}^{\delta_1} T_{m}^{\delta_2} \right)  + \left( \frac{\mu +
\delta_2 - 1}{1-\delta_1} \right) \log \left( W \right) \nonumber \\
& {\hspace{1.5in}} - \left( \frac{\mu}{1-\delta_1} \right) \log
\left( P \right). \label{TvsW}
\end{align}
For example, with $T_{m}W_{d} = 10^{-6}$, $\frac{P}{N_{0}} = 30$
dB, $W = 1$ GHz and a sparsity of $\delta_1 = \delta_2 = 0.5$, the
required minimum signaling duration to obtain first-order
optimality ($\epsilon = 0$, $\mu > 1$) is $T \approx 1$ ms.

Note from (\ref{TvsW}) that smaller $\delta_i$'s imply a slower
scaling of $T$ with $W$. Conversely, for a given $T$ and $W$,
(\ref{TvsW}) can be used to determine the effective value of $\mu$
in (\ref{mu_eps}) as
\begin{equation}
\mu_{\eff} = \frac{\left(1-\delta_1 \right) \log(T/c) +
\left(1-\delta_2 \right) \log(P)}{\log(W/P)}  + \left(1-\delta_2
\right) \label{mueff}
\end{equation}
where $c = \left(T_{m}^{\delta_2} W_{d}^{\delta_1}
\right)^{\frac{1}{1-\delta_1}}$. The effective operational
coherence level can then be determined as $\epsilon_{\eff} =
\frac{ \mu _{\eff}-1}{2}$.

\begin{figure}[hbt!]
\begin{center}
\begin{tabular}{c}
\begin{minipage}{3.45in}
\includegraphics[width=3.4in]{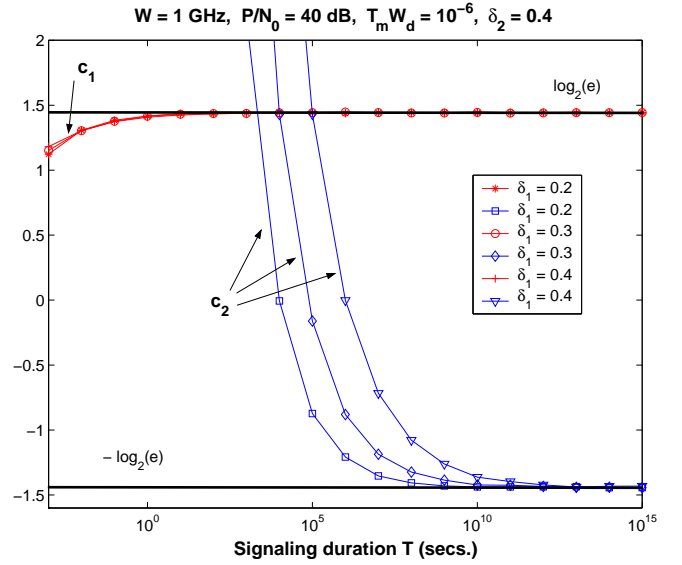}
\end{minipage}
\end{tabular}
\caption{\sl Numerically estimated values of capacity metrics.
Convergence of the coefficients of the $\snr$ and $\snr^{2}$ terms
in capacity as a function of $T$. } \label{fig_c1c2}
\end{center}
\end{figure}

Note that $\mu_{\eff} \rightarrow \infty$ as $T \rightarrow
\infty$ for sparse channels, which implies that any operational
level of coherence can be achieved by simply increasing $T$. This
is due to multipath sparsity in Doppler. This is illustrated in
Fig.~\ref{fig_c1c2}, where we consider the low $\snr$ asymptote of
the coherent capacity in (\ref{coh_bounds}).  The coefficients of
the first- and second-order terms are $\lambda_{1} = \log_{2}(e)$
and $\lambda_{2} = - \log_{2}(e)$, respectively. In
Fig.~\ref{fig_c1c2}, we plot the numerically estimated values
$c_1$ and $c_2$ of $\lambda_1$ and $\lambda_2$, respectively, for
the training-based scheme, which are estimated using Monte-Carlo
simulations and using the optimized lower bound on $I_{tr}$ in
(\ref{I2opt}). For a large enough $T$ such that $\mu_{\eff}
> 1$, the first-order constant $c_{1} \rightarrow \lambda_{1} =
\log_{2}(e)$. Also shown in the figure is the behavior of the
second-order constant and for an even larger value of $T$, we
obtain $c_{2} \rightarrow \lambda_{2} = - \log_{2}(e)$, when
$\mu_{\eff} > 3$.

\begin{figure*}[htb!]
\begin{center}
\begin{tabular}{cc}
\begin{minipage}{3.25in}
\centerline{\includegraphics[width=3.2in]{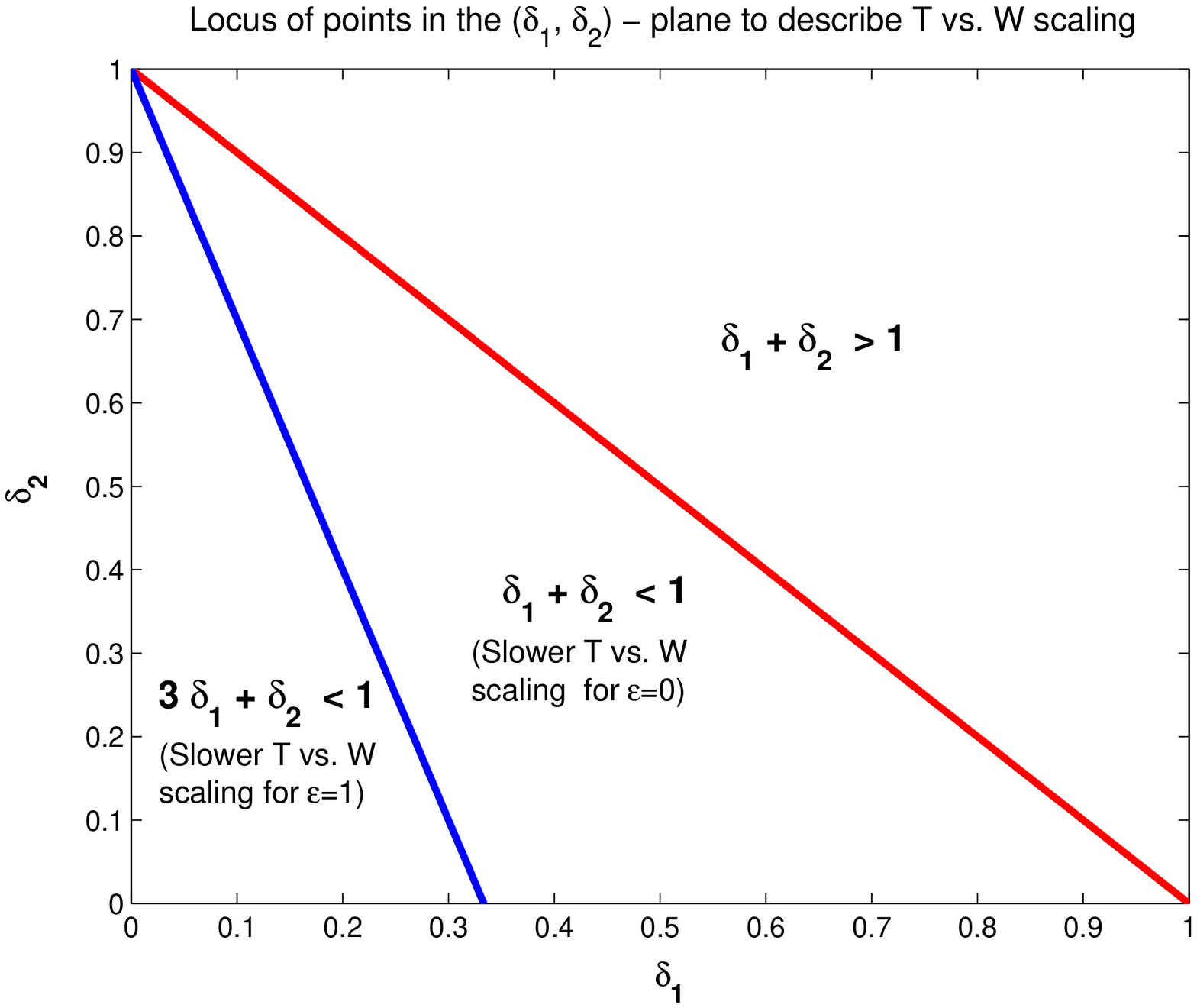}}
\end{minipage}
&
\begin{minipage}{3.25in}
\centerline{\includegraphics[width=3.2in]{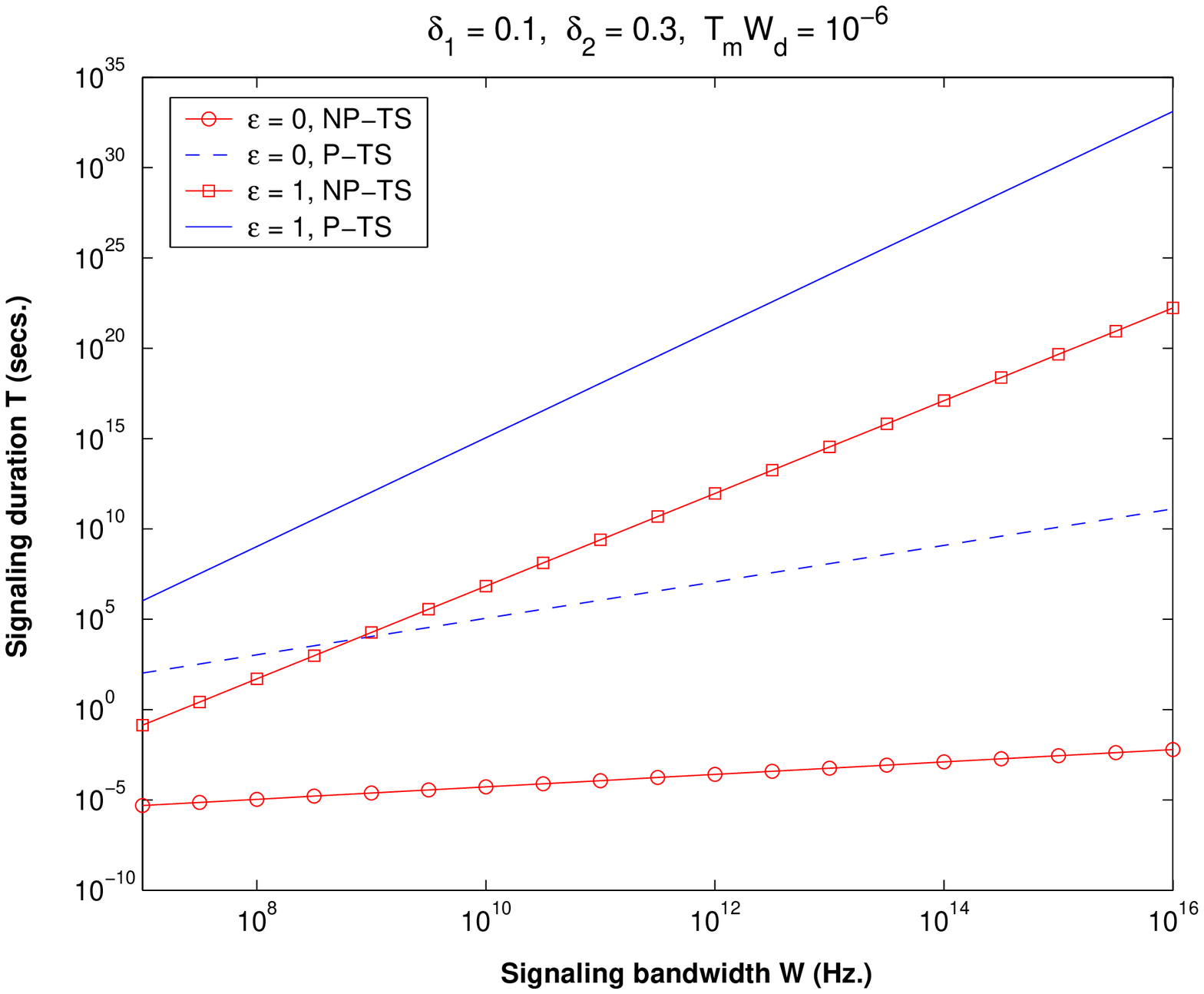}}
\end{minipage} \\ (a) & (b) \\
\begin{minipage}{3.25in}
\centerline{\includegraphics[width=3.2in]{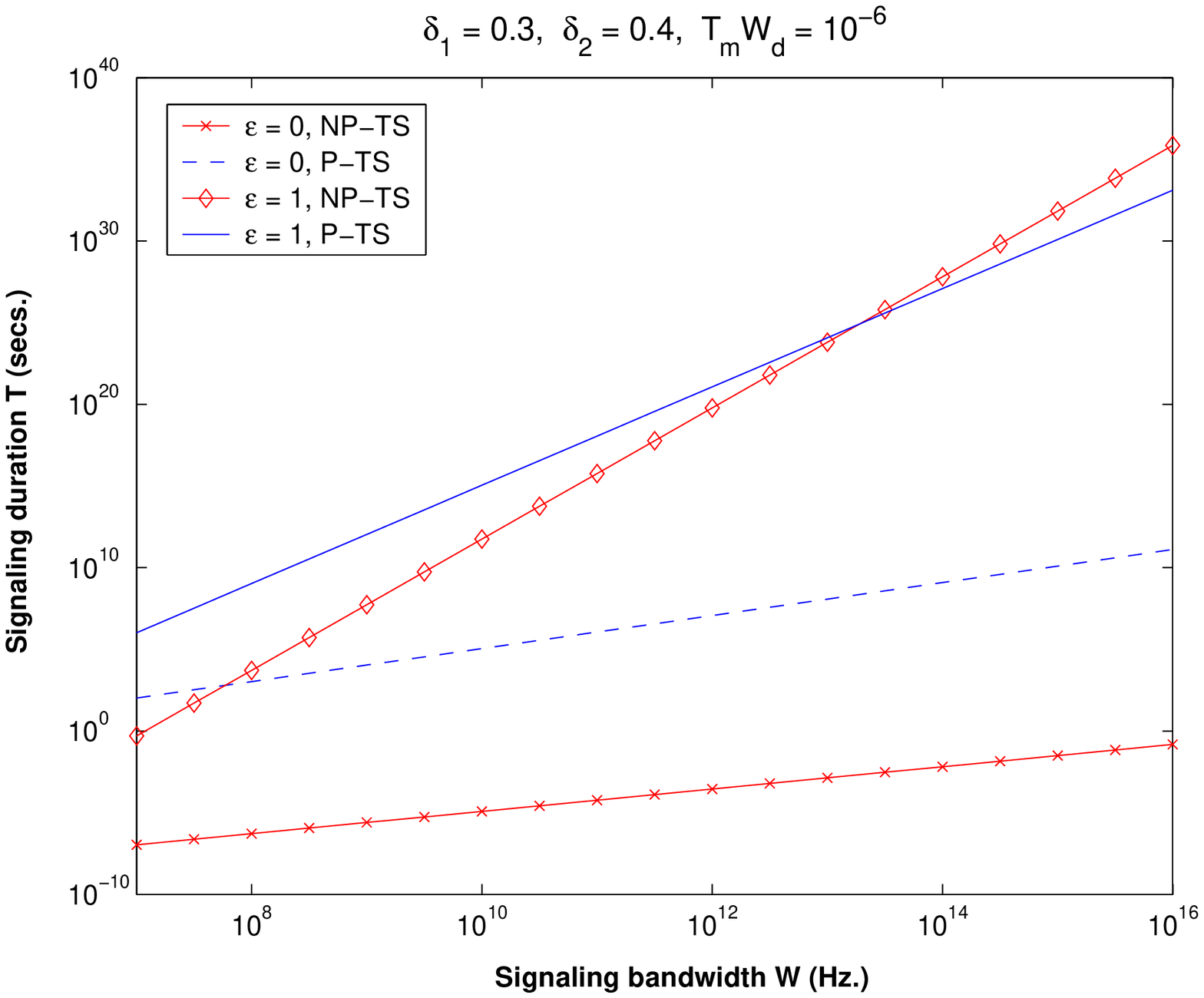}}
\end{minipage}
&
\begin{minipage}{3.25in}
\centerline{\includegraphics[width=3.2in]{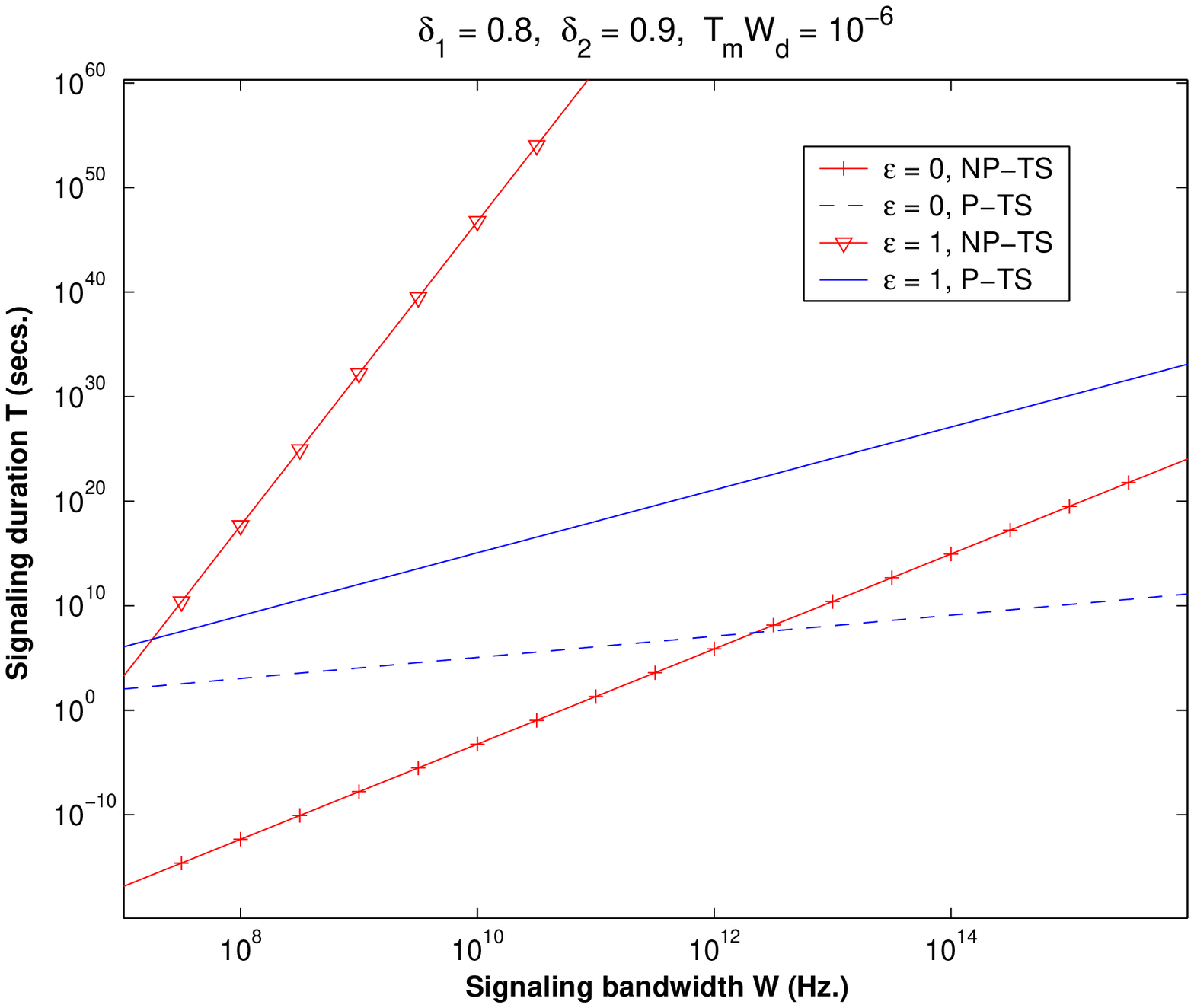}}
\end{minipage} \\ (c) & (d) \\
\end{tabular}
\caption{{\sl (a) Regions in the ($\delta_1, \delta_2$) plane
comparing the required $T$ vs. $W$ scaling in the {\bf NP-TS} and
{\bf P-TS} schemes. Points to the left of the $\delta_1 + \delta_2
= 1$ line represent the favorable region for first-order
optimality ($\epsilon = 0$) of {\bf NP-TS}, illustrated in (c);
for points to the right of this line, {\bf P-TS} yields more
favorable scaling, illustrated in (d). Points to the left of the
$3 \delta_1 + \delta_2 < 1$ line represent the favorable region
for second-order optimality ($\epsilon=1$) of {\bf NP-TS},
illustrated in (b). (b)-(d): $T$ vs. $W$ scaling comparison for
the two schemes for different levels of sparsity. (b) High
sparsity: $\delta_1 = 0.1$ and $\delta_2 = 0.3$. (c) Medium
sparsity: $\delta_1 = 0.3$ and $\delta_2 =  0.4$. (d) Low
sparsity: $\delta_1 = 0.8$ and $\delta_2 = 0.9$.}}
\label{fig_TWscaling}
\end{center}
\vspace{-5mm}
\end{figure*}

\subsection{Peaky versus Non-Peaky Signaling} \label{sec4d} Several
works have emphasized the necessity of signaling schemes that are
peaky in time and/or frequency for achieving wideband capacity in
the non-coherent regime \cite{kennedy,telatar_tse,verdu}. The
motivation behind peaky signaling is that communication takes
place over a smaller set of signaling dimensions, thereby reducing
the effect of channel uncertainty since fewer channel parameters
need to be estimated. However, peaky signaling is practically
infeasible due to peak power constraints. More importantly, the
requirement of peakiness in these works is tied with the implicit
assumption of rich multipath.

When the channel is sparse, the coherence dimension $\nc$
naturally scales with the signal space dimension ($N=TW$) and this
new effect raises the following question: Is peaky signaling still
necessary to achieve capacity in the wideband limit?
Theorem~\ref{thm1} provides the answer: as long as $\mu > 1$,
non-peaky i.i.d. Gaussian input signals are first-order optimal
and with $\mu > 3$, second-order optimality is also satisfied.
While the authors in \cite[Lemma 2]{zheng} (using a non-peaky
training-based communication scheme) obtained exactly the same
conditions on $\mu$, their results are for the scaling of
$T_{coh}$, whereas our scaling result in Theorem~\ref{thm1} is for
$\nc = T_{coh} W_{coh}$. In order to weaken the $T_{coh}$
requirement, the authors in \cite[Lemma 3]{zheng} advocate the use
of peaky training and communication. Furthermore, the
capacity-optimal scheme according to \cite[Theorem 4]{zheng} is a
peaky non-coherent communication scheme in which no explicit
training is performed. Next, we present a detailed discussion on
the scaling laws of $T$ as a function of $W$ to achieve a desired
level of operational coherence. To illustrate the impact of
sparsity, we compare the scaling requirements in this paper with
those in \cite{zheng}.

From (\ref{TvsW}), we note that to achieve an operational coherence
level of $\epsilon$,  $T$ must scale with $W$ as
\begin{equation}
T_{\sparse} \propto W^{\frac{2\epsilon + \delta_2}{1-\delta_1}}
\label{TWscaling}
\end{equation}
where the subscript on $T$ emphasizes that it applies to sparse
channels.  On the other hand, the corresponding scaling on $T$ for
either the peaky or the non-peaky training-based communication
scheme in \cite[Lemma 2 and 3]{zheng}, can be inferred as
\begin{equation}
T_{\rich} \propto \snr^{-(1+2 \epsilon)} \propto W^{1+2 \epsilon}
\label{TWzheng}
\end{equation}
This is because when there is no peakiness, then the minimum
signaling duration is $T = \Tcoh \propto \snr^{-(1+2 \epsilon)}$.
When peaky training and communication is used, $T = L \cdot \Tcoh
\propto \left[\snr^{\epsilon - 1} \right] \cdot \left[\snr^{-3
\epsilon} \right] = \snr^{-(1+2 \epsilon)}$.

Thus, (\ref{TWscaling}) yields a slower (less stringent) scaling
than (\ref{TWzheng}) when
\begin{equation}
\frac{2\epsilon + \delta_2}{1-\delta_1}  <  1+2 \epsilon
\Longleftrightarrow   \left( 1+2 \epsilon \right) \delta_1 + \delta_2
< 1. \label{deltailocus}
\end{equation}
The locus of points in the ($\delta_1,\delta_2$) plane represented
in (\ref{deltailocus}) defines the set of channel sparsity values
for which we obtain a slower scaling requirement. This is
pictorially represented in Fig.~\ref{fig_TWscaling}(a) for the
special cases of $\epsilon=0$ (first-order optimality) and
$\epsilon=1$ (second-order optimality). 

Figs.~\ref{fig_TWscaling}(b)-(d) illustrate the required scaling
of $T$ with $W$ for different levels of channel sparsity. In all
figures, the non-peaky training-based scheme in our framework is
denoted by \textbf{NP-TS}, whereas the peaky training scheme in
\cite{zheng} is denoted by \textbf{P-TS}. The signaling duration
requirements for \textbf{P-TS} are independent of channel sparsity
and are given by
\begin{equation}
T_{p-ts,1} \propto W \ \ , \ \ T_{p-ts,2} \propto W^3 \label{T_rich}
\end{equation}
where the subscripts ``1'' and ``2'' reflect the requirements for
first- and second-order optimality, respectively.
Fig.~\ref{fig_TWscaling}(b) compares the scaling requirements for
the sparsest channel: $\delta_1 = 0.1$ and $\delta_2 = 0.3$ so that
$3 \delta_1 + \delta_2 < 1$. In this case, the scaling requirements
for \textbf{NP-TS} are: \begin{equation} T_{np-ts,1} \propto W^{1/3}
< W \ \ , \ \ T_{np-ts,2} \propto W^{2.3/0.9} < W^3
\end{equation} which are less stringent that (\ref{T_rich}) for both
first- and second-order optimality. Fig.~\ref{fig_TWscaling}(b)
corresponds to a medium sparse channel: $\delta_1 = 0.3$ and
$\delta_2 = 0.4$. In this case, the scaling requirements for
\textbf{NP-TS} are
\begin{equation} T_{np-ts,1} \propto W^{0.4/0.7} < W \ \ , \ \
T_{np-ts,2} \propto W^{2.4/0.7} > W^3 \end{equation} which are less
stringent than (\ref{T_rich}) for first-order optimality but more
stringent for second-order optimality. Fig.~\ref{fig_TWscaling}(c)
represents the least sparse channel: $\delta_1 = 0.8$ and $\delta_2
= 0.9$ so that $\delta_1 + \delta_2
> 1$. In this case, the scaling requirements for \textbf{NP-TS} are
\begin{equation} T_{np-ts,1} \propto W^{0.9/0.2} > W  \ \ , \ \
T_{np-ts,2} \propto W^{2.9/0.2} > W^3 \end{equation} which are more
stringent than (\ref{T_rich}) for both first- and second-order
optimality.

\subsection{Rich versus Sparse Multipath: The Extreme Cases}
\label{sec4e} We now discuss the two extreme scenarios of rich and
sparse multipath, i.e, $\delta_i \rightarrow 0 \; \text{or} \;
1,\hsp i=1,2$. The canonical scaling relationship in
(\ref{TWP_locus}) between $T$ and $W$ (ignoring constants) is
\begin{equation}
T \propto W^{\frac{\mu + \delta_2 - 1}{1-\delta_1}}. \label{TW_np}
\end{equation}
As either $\delta_1$ or $\delta_2$ or both tend to zero, we have a
very sparse channel in which any desired value of $\mu$ can be
obtained with relatively small values of $T$ by following
(\ref{TW_np}).

When $\delta_2 \rightarrow 1$, the conditions on $T$ in
(\ref{TW_np}) grow more stringent in order to attain a desired
$\mu$. When $\delta_2 = 1$, $W_{coh}$ is a constant and the
requirements on $\nc$ can be attained only through $T_{coh}$
scaling with increasing $T$. In particular, the conditions on $T$
in (\ref{TW_np}) become
\begin{equation}
T \propto W^{\frac{\mu}{1-\delta_1}}. \label{TW_delta1}
\end{equation}
As $\delta_1 \rightarrow 1$, the conditions on $T$ to attain a
desired $\mu$ become more stringent. When $\delta_1 = 1$, we have
a constant $T_{coh}$ and from a scaling perspective, $\nc =
W_{coh} \propto W^{1-\delta_2} = \frac{1}{\snr^{1-\delta_2}}$.
Thus the attained value of $\mu$ is $\mu = 1-\delta_2 \leq 1$, and
even first-order optimality cannot be obtained.

This issue can be resolved by considering peaky signaling schemes,
that also help offset the large $T$ requirements when $\delta_1$
and/or $\delta_2$ is close to $1$. We model peaky signaling by
assuming that a subset of the time-frequency coherence subspaces
in each codeword (Fig.~\ref{fig:del_dopp}(b)) are used for
training and communication and no information is sent in the
remaining subspaces. We model peakiness similar to \cite{zheng}
and define
\begin{equation}
\zeta = \snr^{\gamma}, \hsp \hsp \gamma > 0 \label{peaky}
\end{equation}
as the fraction of signal space dimensions which are used for
communication. The effect of peakiness is captured through the
parameter $\gamma$. More specifically, the peakiness ratio
($\sf{PR}$) between peaky and non-peaky signaling given by
$\sf{PR} = \frac{\snr^{'}}{\snr} = \snr^{-\gamma} \rightarrow
\infty$ as $\snr \rightarrow 0$ since $\gamma > 0$. It is clear
that $\gamma < 1$, since the energy per transmit symbol equals
\begin{equation}
\snr^{'} = \frac{\snr}{\snr^{\gamma}} = \snr^{1-\gamma}
\label{snrp}
\end{equation}
and $\snr^{'} \geq 1$ when $\gamma \geq 1$ and we are no longer in
the low $\snr$ regime. The following result captures the impact of
peakiness on the average mutual information of the training-based
scheme.
\begin{prp}
\label{lem3} The peaky training-based scheme achieves
\begin{equation}
I_{tr}^{p}(\snr) \geq \log_2(e) \cdot \left[ \snr - \ord \left(
\snr^{\frac{1+\mu}{2}} \right) \right] \label{cap_tr_peaky}
\end{equation}
if
$\nc = 1/{\snr^{\mu - \gamma}}.$ 
\end{prp}
\begin{proof}
The average mutual information with a peaky input equals
\begin{equation}
I_{tr}^{p}(\snr) = \zeta \hsp I_{tr}(\snr^{'}) = \snr^{\gamma}
\hsp I_{tr}(\snr^{'}) \label{Ipeaky}
\end{equation}
where $I_{tr}(\snr^{'})$ is the average mutual information
achievable with the non-peaky scheme, as in (\ref{cap_tr}) of
Theorem~\ref{thm1}. Therefore, if
\begin{equation*}
\nc = \frac{1}{\snr^{\mu-\gamma}} =
\frac{1}{\left(\snr^{'}\right)^{\frac{\mu-\gamma}{1-\gamma}}}
\end{equation*}
then, using (\ref{Ipeaky}) and (\ref{cap_tr}), we have
\begin{align}
& {\hspace{-0.1in}} I_{tr}^{p}(\snr)  \nonumber \\
&  \geq \log_2(e) \cdot \tsty{ \snr^{\gamma} }
\cdot \left[ \tsty{ \snr^{'}} - \ord \left( \left( \tsty{
\snr^{'}} \right)^{\frac{1+\frac{\mu-\gamma}{1-\gamma}}{2}}
\right) \right] \nonumber \\
& \stackrel{\mathit{(a)}}{=}  \log_2(e) \cdot \tsty{
\snr^{\gamma} } \cdot \left[ \tsty{ \snr^{1-\gamma}} - \ord \left(
\tsty{ \snr^{\frac{1+\mu-2
\gamma}{2}} } \right) \right] \nonumber \\
& =  \log_2(e) \cdot \left[ \snr - \ord \left(
\snr^{\frac{1+\mu}{2}} \right) \right] 
\end{align}
where (a) follows from (\ref{snrp}). This proves the proposition.
\end{proof}
Thus the advantage of using a peaky input manifests itself in
reducing the required $\snr$ exponent of the coherence dimension,
$\nc$. That is, the effective $\mu$ reduces to $\mu_{\peaky} = \mu
- \gamma$.
Using the result of Proposition~\ref{lem3}, we now revisit the scaling law
in (\ref{TWP_locus}). As a consequence of the condition $\nc = 1/{\snr^{\mu - \gamma}}$,
we
obtain a slower (relaxed) scaling of $T$ as a function of $W$ to
achieve a desired value of $\mu$
\begin{equation}
T \propto W^{\frac{\mu + \delta_2 - 1 - \gamma }{1-\delta_1}}.
\label{TvsW_peaky}
\end{equation}
For any $0 < \delta_1,\delta_2 <1$, the rate at which $T$ scales
with $W$ can now be controlled through the peakiness parameter
$\gamma$, especially when $\delta_i \rightarrow 1$. More
importantly, when $\delta_1 = 1$, we have $\nc = W_{coh} =
\frac{1}{\snr^{1-\delta_2}}$ and therefore we can satisfy the
condition $\nc = 1/{\snr^{\mu - \gamma}}$ as long as
\begin{equation}
\gamma \geq \mu + \delta_2 -1. \label{gamma_delta1zero}
\end{equation}
Note that while we can obtain first-order optimality in this case,
and necessarily through peaky signaling, second-order optimality
is not feasible since it requires $\gamma \geq (2+\delta_2) > 1$.
When $\delta_2 = 1$, peakiness is not necessary, but the scaling
requirements on $T$ can be relaxed from (\ref{TW_delta1}) to
\begin{equation}
T \propto W^{\frac{\mu - \gamma}{1-\delta_1}}.
\label{TW_delta1_peaky}
\end{equation}

\subsection{Arbitrary Sub-linear Scaling Laws}
\label{sec4f} We modeled sparsity in delay and Doppler by
restricting our attention to the power-law scaling in
(\ref{sparse}). We now show that the results in this paper hold
true for \emph{any} sub-linear scaling in the DoF. Since sparsity
in delay/Doppler implies that $W_{coh}$ and $T_{coh}$ scale
(sub-linearly) with $W$ and $T$ respectively, we assume a general
scaling law for these quantities. Let
\begin{equation}
W_{coh} = \fone\left( W \right), \hsp \hsp \hsp T_{coh} = \ftwo
\left( T \right) \label{TcohWcoh}
\end{equation}
\begin{equation}
\Longrightarrow \nc = T_{coh} W_{coh} = \fone(W) \ftwo(T)
\label{nc_scaling}
\end{equation}
where $\fone$ and $\ftwo$ are strictly increasing, arbitrary
sub-linear functions of $W$ and $T$ respectively. That is,
$\fone(W) \sim \littleo(W)$ and $\ftwo(T) \sim \littleo(T)$. Note
that the definition in (\ref{TcohWcoh}) implies that $D_{W} =
\frac{W}{\fone(W)} \sim \littleo(W)$ and $D_{T} =
\frac{T}{\ftwo(T)} = \littleo(T)$. We also assume
\begin{equation}
T = \fthr(W) \label{TWf3}
\end{equation}
where $\fthr$ reflects the scaling of $T$ with $W$, necessary to
obtain a desired value of $\mu$. Given $\fone$ and $\ftwo$, our
focus here is to find a suitable $\fthr$ so that a desired value
of $\mu$ can be obtained.

A key observation from Theorem~\ref{thm1} is that it provides
necessary and sufficient conditions for first- and second-order
optimality that are \emph{independent} of the power-law scaling
assumptions in (\ref{sparse}). Recall that with $\nc =
\frac{1}{\snr^{\mu}}$, the condition for first-order optimality is
$\mu > 1$ and that for second-order optimality is $\mu > 3$.
Defining a new parameter $E_{d} = \nc \snr = \snr^{1-\mu}$, which
has the physical interpretation of the transmit energy per DoF, we
have in the limit of $\snr \rightarrow 0$,  $E_{d} \rightarrow
\infty$ as $\ord \left(\frac{1}{\snr^{\mu - 1}} \right)$ with $\mu
> 1$ and $\mu > 3$ for first- and second-order optimality,
respectively. Using (\ref{Ncoh}) and (\ref{TWf3}), we have
\begin{eqnarray}
E_{d} = \nc \snr & = & \fone(W) \ftwo(T) \snr \nonumber \\
& = & \fone(W) \ftwo \left( \fthr(W) \right) \snr \nonumber \\
& = & \fone(W) \gone(W) \snr \label{Ed_snr}
\end{eqnarray}
where we have defined $\gone(x) = (\ftwo \circ \fthr)(x)$. We also
provide the following definition that is used in the subsequent
theorem.
\begin{defn}
\label{defn2} For any two functions $f$ and $g$, we define
\begin{equation}
f(x) \sim w(g(x)) \hsp \hsp \Longleftrightarrow \hsp \hsp \lim
\limits_{x \rightarrow \infty} \left| \frac{f(x)}{g(x)} \right| =
\infty. \label{fgdef}
\end{equation}
\end{defn}
{\vspace{0.1in}}
\begin{thm}
\label{thm3} For the coherence scaling laws in (\ref{TcohWcoh})
and (\ref{nc_scaling}), a necessary and sufficient condition to
obtain a desired value of $\mu$ is given by
$\fone(x) \gone(x) \sim w(x^{\mu}).$ 
\end{thm}
{\vspace{0.1in}}
\begin{proof}
Using (\ref{Ed_snr}) and noting that $\snr = \frac{P}{W}$, we have
\begin{eqnarray}
\nc \snr  =  \fone \left( \tsty{ \frac{1}{\snr} } \right) \gone
\left( \tsty{ \frac{1}{\snr} } \right) \snr 
 = \frac{\fone \left( x \right) \gone \left( x \right)}{x}.
\nonumber
\end{eqnarray}
Therefore, to obtain a specific $\mu$, we require
\begin{align}
& \nc \snr  =  \ord
\left(\frac{1}{\snr^{\mu-1}}\right) 
\Longleftrightarrow \frac{\fone \left( x \right) \gone \left( x
\right)}{x} =  \ord \left(x^{\mu-1} \right) \nonumber \\
& {\hspace{1in}}
\Longleftrightarrow \fone(x) \gone(x)  \sim w(x^{\mu}).
\end{align}
\end{proof}
Note that the conditions for first- and second-order optimality
are $\fone(x) \gone(x) \sim w(x)$ and $\fone(x) \gone(x) \sim
w(x^{3})$, respectively.
\begin{cor}
\label{cor1} For given $\fone$ and $\ftwo$, the conditions of
Theorem~\ref{thm3} are satisfied by choosing $\fthr(x) =
\ftwo^{-1} \left( \frac{x^{\mu}}{\fone(x)} \right)$.
\end{cor}
\begin{rem}
\label{rem3} The conditions of Theorem~\ref{thm3} are satisfied
under the power-law scaling assumptions in (\ref{Ncoh}) and the
$T$ vs. $W$ scaling relationship in (\ref{TWP_locus}). We have
$\fone(x) = x^{1-\delta_2}$, $\ftwo(x) = x^{1-\delta_1}$,
$\fthr(x) = x^{\frac{\mu-1+\delta_2}{1-\delta_1}}$ and it follows
that $\fone(x) \gone(x) = x^{\mu}$.
\end{rem}

\subsection{Comments on Channel Modeling}
\label{sec4g} A couple of comments on the channel model used in
this paper are warranted. First, the block fading channel model in
the STF domain used in this paper is an idealization of the
effects of multipath sparsity in delay-Doppler. The idealized
model was used to facilitate capacity analysis by relating the
sub-linear scaling in the channel DoF in delay-Doppler to the
scaling in the time-frequency coherence dimension under STF
signaling. While the actual channel in the STF domain would
exhibit more complex characteristics, the block fading
idealization does capture the essence of multipath sparsity from
the viewpoint of DoF scaling, which is the most important channel
property in the context of channel capacity in the limit of large
signal space dimension.

Second, throughout this work, we assume a simplistic Gaussian
model for small-scale fading. However, evidence from measurement
campaigns suggests ``specular'' statistics for the channel
coefficients and some channel
measurements~\cite{molisch,chiachin1} indicate that Nakagami or
log-normal distributions may be a more accurate fit for the
small-scale fading in the wideband regime. While this issue is not
addressed in this paper, our assumption of Gaussian statistics
permits closed-form analysis and we suspect that the implications
of multipath sparsity would hold under such statistics as well.

\section{Conclusions}
\label{sec5} We have investigated the ergodic capacity of sparse
multipath channels in the ultrawideband regime. Motivated by
recent measurement campaigns, we have introduced a model for
sparse multipath channels that captures the effect of multipath
sparsity on the statistically independent DoF in the channel via
the notion of resolvable paths in delay and Doppler. The workhorse
of our analysis is the use of orthogonal STF signaling that
approximately diagonalizes underspread channels and naturally
relates multipath sparsity in delay-Doppler to coherence in time
and frequency. In particular, we proposed a simple block-fading
model for sparse channels in the STF domain that captures the
sub-linear scaling of the channel DoF with signal space
dimensions.

Our work builds on recent results on ergodic capacity in the
wideband regime to study the impact of multipath sparsity on
bridging the gap between coherent and non-coherent regimes. The
most significant implication of multipath sparsity is that the
requirements on coherence time, $\Tcoh$, in existing works
\cite{zheng} are naturally replaced by requirements on the
time-frequency coherence dimension, $\nc = \Tcoh \Wcoh$. As a
result the requirements on channel coherence are shared between
time and frequency thereby leading to significantly reduced
coherence time requirements to attain a desired level of
coherence. Our results reveal how any desired operational
coherence can be achieved by scaling the signaling parameters --
signaling duration $T$, bandwidth $W$ and transmit power $P$ -- in
an appropriate fashion. We also discussed the usefulness of peaky
signaling schemes for reducing coherence requirements and the role
played by channel sparsity in relaxing peakiness requirements.

There are many interesting directions for future work. First, it
would be useful to refine the results in this paper via more
accurate modeling of sparsity in the time-frequency domain (as
opposed to the block fading model). Second, studying the impact of
non-Gaussian statistics of channel coefficients would also be
useful. Third, while ergodic capacity is achieved by coding over
long signaling durations, in practical settings with strict delay
constraints, it is important to investigate more relevant metrics,
like outage capacity \cite{ozarow}. An important and related
performance metric is reliability (in terms of error exponents)
\cite{gallager_book}. We are currently investigating the impact of
multipath sparsity on outage capacity and reliability. In this
context, we recently reported a new fundamental {\em learnability
versus diversity tradeoff} in sparse channels that governs the
impact of sparsity on reliability and error probability
\cite{allerton_06}. Another interesting aspect to study is the
impact of feedback on achievable rates
\cite{borade},\cite{manish}. Finally, we note that sparse channel
models arise in other scenarios as well, such as underwater
acoustic channels (see e.g.,~\cite{carbonelli}). Thus the
implications of this work may be applicable in such situations as
well.

\appendix

\subsection{Proof of Proposition \ref{lowerbound}}
\label{append:app1} As is well known, the coherent capacity
expression can be computed in closed-form using standard integral
formulas. For this, we use the following
fact~\cite[4.337(1), pp.\ 574]{gradshteyn}:
\begin{align}
& {\hspace{-0.05in}}
\int_0^{\infty} \log_e(a + x) e^{-bx} \ud x = \frac{1}{b} \left[
\log_e(a) + e^{ab} \int_{ab}^{\infty} \frac{e^{-t} \ud t}{t}
\right]. \label{explog_int}
\end{align}
Particularizing (\ref{explog_int}) to ${\bEe}\left[ \log_2 \left( 1 +
{\snr} \left| h \right| ^2 \right)  \right]$ by a transformation of random
variables of the form ${\mathrm{Re}}(h) = r\cos(\theta),
{\mathrm{Im}}(h) = r\sin(\theta)$ results in
\begin{eqnarray}
C_{coh}(\snr) 
& = &  e^{\frac{1}{\snr}} \int_{\frac{1}{\snr}}^{\infty}
\frac{e^{-t}}{t} \ud t.
\end{eqnarray}
We can then bound $C_{coh}(\snr)$ using~\cite[5.1.20,
pp.229]{abramowitz} as
\begin{equation}
\label{boundsforexp} \frac{1}{2} \log_e \left( 1 + 2\snr \right)
\leq e^{ \frac{1}{\snr} } \int_{ \frac{1}{\snr} }^{\infty}
\frac{e^{-t} \ud t}{t} \leq \log_e \left( 1 + \snr \right).
\end{equation}
The upper bound of the proposition follows from a combination of
Jensen's inequality and the monotonicity of
$\log_e(1+x)-x+\frac{bx^2}{2}$ under the imposed constraints on
$b$. The lower bound follows via a Taylor's series truncation. The
tightness of the lower bound at low $\snr$ follows from the
asymptotic (in $\frac{1}{\snr}$) expansion of the exponential
integral~\cite[5.1.51, pp.\ 231]{abramowitz}.
\endproof

\subsection{Proof of Lemma \ref{lem1}}
\label{append:app2} We begin with the vectorized system equation
for the communication component of the scheme (described in
(\ref{comm_split}))
\begin{equation}
\by = \bH \bx + \bw = \widehat{\bH} \bx + \mathbf{\Delta} \bx +
\bw. \label{comm_matrix}
\end{equation}
Here, we have represented the $(\nc-1)D$-dimensional communication
sub-channel of the diagonal channel in (\ref{disc_channel}) by
$\bH$ for simplicity. $\widehat{\bH }$ is the
$(\nc-1)D$-dimensional diagonal matrix of channel estimates and
$\mathbf{\Delta}$ is the estimation error matrix, $\mathbf{\Delta}
= {\bH} - \widehat{\bH }$. Lumping the estimation error along with
the additive noise and optimizing over the set of input covariance
matrices $\bQ$ that satisfy ${\mathrm{Tr}} \left( {\bQ} \right) =
(1-\eta)\hsp TP$, a lower bound to $I_{tr}$ is
achieved~\cite{medard} as follows:
\begin{align}
& {\hspace{-0.07in}} I_{tr}  \geq  
\sup_{\bQ} \cdot \frac{ \tsty{
 {\bEe} \left[ \log_2 \det \left( \bI + \widehat{\bH } {\bQ}
 \widehat{\bH }^H \left( \bI + \Sigma_{ {\mathbf{\Delta}}
 {\mathbf{x}} } \right)^{-1} \right) \right] }}{\nc D} \label{lb1}
\end{align}
where $\bI$ denotes the $(\nc-1)D$ dimensional identity matrix. We
use a zero-mean Gaussian input with covariance matrix $\bQ =
\frac{ {\mathrm{Tr}} \left( {\bQ} \right) } { (\nc-1)D} \bI $.
With this choice, note that $\Sigma_{{ \mathbf{\Delta}}
{\mathbf{x}} } = \bEe_{\bH, \hsppp \bx} \left[ {\mathbf{\Delta}}
{\mathbf{x}} {\mathbf{x}}^H {\mathbf{\Delta}}^H \right] =
\bEe_{\bH} \left[ {\mathbf{\Delta}} \bQ {\mathbf{\Delta}}^H
\right] = \frac{1}{1 + \etr} \cdot \frac{ {\mathrm{Tr}} \left(
{\bQ} \right) }
{ (\nc-1)D} \hsp \bI $ since 
${h}_i$ are identically distributed. Thus, we have
\begin{eqnarray}
I_{tr}  & \geq &  \frac{1}{\nc D} \cdot {\bEe} \left[ \log_2 \det
\left( \bI + \beta
 \widehat{\bH }  \widehat{\bH }^H \right) \right]
 \nonumber \\
  & = & \left(\frac{\nc-1}{\nc D}\right) \cdot
  \sum_{i=1}^D {\bEe} \left[ \log_2 \left( 1 + \beta
  \left| \widehat { {{h}} }_i\right| ^2 \right)  \right]
  \nonumber \\
& \stackrel{\mathit{(a)}}{=} & \left( 1  - \frac{1}{\nc} \right )
\cdot {\bEe} \left[ \log_2 \left( 1 + \beta   \left| \widehat
{h}\right| ^2 \right)  \right] \label{eqnrefer1}
\end{eqnarray}
where $\beta$ is as in (\ref{beta}) and (a) follows because the
random variables $\{\widehat{h}_{i}\}$ are i.i.d. Furthermore, it
can be shown that the $\widehat{h}_i$'s are zero-mean with ${\bEe}
[ | \widehat{h}_i | ^2 ] = {\bEe} [| \widehat{h} | ^2 ] = \sigma^{2}$ as in
(\ref{sigmasq}). 
We now compute the expectation in (\ref{eqnrefer1}) in
closed-form. For this, we use
(\ref{explog_int})~\cite[4.337(1), pp.\ 574]{gradshteyn}.
Particularizing (\ref{explog_int}) to ${\bEe} \left[ \log_2 \left(
1 + \beta \left| \widehat{h} \right| ^2 \right)  \right]$ by a
transformation of random variables of the form
${\mathrm{Re}}(\widehat{h}) = r\cos(\theta),
{\mathrm{Im}}(\widehat{h}) =
r\sin(\theta)$ results in 
\begin{eqnarray}
I_{tr} &\geq & 
\left( 1 - \frac{1}{\nc} \right ) \cdot \log_2(e) \cdot e^{
\frac{1}{\beta \sigma^2} } \int_{ \frac{1}{\beta \sigma^2}
}^{\infty} \frac{e^{-t} \ud t}{t} \label{closedform}
\end{eqnarray}
While (\ref{closedform}) provides a closed-form lower bound for
$I_{tr}$,
we need a more tractable estimate for the same. 
For this, we use
(\ref{boundsforexp})~\cite[5.1.20,pp.229]{abramowitz}. Thus
$I_{tr}$ can be further lower bounded as
\begin{eqnarray}
I_{tr} & \geq  & \widehat{I}_{tr} \triangleq  \frac{1}{2} \log_e
\left( 1 + 2\beta \sigma^2 \right). \label{estdbound}
\end{eqnarray}
This completes the proof of the lemma.
\endproof

\subsection{Proof of Lemma \ref{lem2}}
\label{append:app3} Since $\log(\cdot)$ is a monotonically
increasing function, the tightest lower bound to $I_{tr}$ is
obtained by maximizing $K(\eta)$. A tedious, but straightforward,
computation shows that for any $a,b > 0$, the function
$f(\eta,a,b)$ defined on $\eta \in [0,1]$ as
\begin{eqnarray}
f(\eta,a,b)  = \frac{\eta (1-\eta)}{ a + b(1 - 2\eta + \eta N_c) }
\end{eqnarray}
is concave as a function of $\eta$. Now note that $K(\eta) = N_c^2
\hspp \snr^2 \hspp f(\eta, N_c - 1, N_c \snr)$. Thus $K(\eta)$ is
maximized by setting its first derivative to zero.

It is easy to check that the $\eta$ that is sought is a root of
the quadratic
\begin{align}
& \eta^2 \left(\nc \snr (\nc-2) \right) +2 \eta \left( \nc \snr +
(\nc-1) \right) \nonumber \\
& {\hspace{0.6in}} - \left( \nc \snr + (\nc-1) \right) = 0 \nonumber
\end{align}
and is precisely $\etas$ as in (\ref{etaopt}). Using this value of
$\etas$ yields the optimal $K^{*}$ as in (\ref{Kopt}). Thus the
lemma has been established.
\endproof

\subsection{Proof of Theorem \ref{thm1}}
\label{append:app4}

Substituting $N_c = \frac{1}{\snr^{\mu}}$ in (\ref{Kopt}), we have
\begin{eqnarray}
K^{*} & = & K_1 K_2  \ , \  K_1  = \tsty{ \frac{\snr^{\mu} \hsp
\left( \snr +  1 - \snr^{\mu} \right)   }
{  \left( 1 - 2\snr^{\mu} \right)^2} } \nonumber \\
K_2 & = & \tsty{ \left[ \sqrt{ 1 + \frac{\snr^{1-\mu} \hsp
\left(1- 2\snr^{\mu}\right) }   {\snr  + 1 -\snr^{\mu} } } - 1
\right]^2 }. \label{K_defn}
\end{eqnarray}
We study the low $\snr$ asymptotics of $K$ for the following four
cases -- Case 1: $\mu = 1$, Case 2: $\mu \in (1,3)$, Case 3: $\mu
\geq 3$ and Case 4: $\mu < 1$.

Case 1: It is not difficult to check that
\begin{equation*}
K_1  =  \tsty{ \snr + \ord(\snr^2) } 
\end{equation*}
\begin{equation*}
K_2  =  \tsty{ \left( \sqrt {2 + \ord(\snr) + \ord(\snr^2) } - 1
\right)^2 = \ord(1) }.
\end{equation*}
Using the above relationships in (\ref{I2opt}), we see that the
coefficient of the $\snr$-term in the low $\snr$ expansion of
$\widehat{I}_{tr}$ is {\emph{strictly}} smaller than $\log_2(e)$.
Thus, first-order optimality fails.

Case 2: When $\mu \in (1,3)$, we have
\begin{eqnarray}
\label{c2} K_1 &=& \tsty{ \snr^{\mu} \sum_{i=\{0,1\}}
\sum_{j=0}^{j=\infty}
\ord \left( \snr^{i + j\mu} \right) } \\
K_2 & = & \tsty{ \frac{1}{\snr^{\mu-1} }} \Bigg[ \tsty{ 1 + 2 \hsp
\snr^{\mu-1} - \snr -2 \hsp \snr^{ \frac{\mu-1}{2} } }   \nonumber \\
& & {\hspace{0.1in}} - \tsty{ \snr^{ \frac{3\mu-3}{2} } + \left( \frac{1}{2}
\right)^2 \snr^{2 \mu -2} } \Bigg]   \label{c3}
\end{eqnarray}
which implies that one of the $\snr^{\mu}$, $\snr^{\frac{\mu
+1}{2}  }$, $\snr^{\frac{3\mu -1}{2}  }$, $\snr^{2\mu -1  }$ terms
in $K$ leads to failure of second-order optimality condition. In
particular, the coefficient of the $\snr^{1+\epsilon}$ term in
(\ref{epslessthanone}) is obtained from the coefficient of the
$\snr^{\frac{\mu-1}{2}}$ term within the parenthesis in
(\ref{c3}). However, we get exact first-order optimality in this
case.

Case 3: When $\mu \geq 3$, $K_1$ and $K_2$ are given by (\ref{c2})
and (\ref{c3}), respectively and every vanishing term is of the
form $\snr$ or $\snr^{\nu}$ for some $\nu \geq 2$. When $\mu = 3$,
we note that the contribution to the coefficient of the $\snr^{2}$
term can be obtained from (\ref{c2}), (\ref{c3}) and equals $ -3
$. When $\mu > 3$, it is easy to see that we get exact
second-order optimality. Thus a low $\snr$ expansion of
$\widehat{I}_{tr}$ in the form we seek is achievable.

Case 4: When $\mu < 1$, $K_1$ is given by the same relationship as
in (\ref{c2}). But for $K_2$ we have
\begin{eqnarray}
K_2 = \tsty{ \left(  \frac{1}{2} \hsp \snr^{1-\mu} \hsp
\sum_{i=0}^{\infty} \sum_{j=0 }^{\infty} \ord \left( \snr^{i +
j\mu} \right) \right)^2 }.
\end{eqnarray}
This results in the failure of the first-order optimality
condition since the largest power of $\snr$ in the Taylor's series
expansion of $\widehat{I}_{tr}$ is $\snr^{2-\mu}$.
\endproof

\subsection{Proof of Theorem \ref{thm2}}
\label{append:app5} We follow the same technique as in Theorem
\ref{thm1}. We rewrite the expression for $\etas$ in
(\ref{etaopt}) (using $\nc = \frac{1}{\snr^{\mu}}$) as
\begin{eqnarray}
\etas & = & \eta_{1} \eta_{2}, \hsp \hsp \hsp \hsp \eta_{1} =
\tsty{ \frac{\snr^{\mu -1} \left(\snr + 1 - \snr^{\mu}
\right)}{\left( 1 - 2
\snr^{\mu} \right)} } \nonumber \\
\eta_{2} & = & \tsty{ \left[ \sqrt{ 1 + \frac{\snr^{1-\mu} \hsp
\left(1- 2\snr^{\mu}\right) }   {\snr  + 1 - \snr^{\mu} } } - 1
\right] }. \label{eta_def}
\end{eqnarray}
To characterize the behavior of $\mseq = \frac{1}{1+\etr}$, we
analyze $\etr = \etas \nc \snr = \snr^{1-\mu} \eta_{1} \eta_{2}$.
We consider the asymptotics in either of the following two
scenarios: (i) fixed $\mu$ and $\snr \rightarrow 0$ (as would be
the case if we increase $W$ and scale $T$ appropriately, according
to (\ref{TWP_locus})) (ii) fixed low $\snr$ ($\ll 1$) and
increasing $\mu$ (for large but fixed $W$ and increasing $T$). The
analysis is done over the following three cases: Case 1: $\mu <
1$, Case 2: $\mu = 1$ and Case 3: $\mu > 1$.

Case 1: When $\mu < 1$, we have
\begin{eqnarray}
\eta_1 &=& \tsty{\snr^{\mu - 1}} \sum_{i=\{0,1\}} \sum_{j=0
}^{j=\infty}
\tsty{\ord \left( \snr^{i + j\mu} \right)} \label{eta1case1} \\
\eta_2 & = &  
\frac{1}{2} \hsp \snr^{1-\mu} \hsp \sum_{i=0}^{\infty}
\sum_{j=0}^{\infty} \ord \left( \snr^{i + j\mu} \right). 
\label{case1}
\end{eqnarray}
This leads to
\begin{eqnarray}
\etas & = & \tsty{ \frac{1}{2} + \sum_{i=0}^{\infty}
\sum_{j=2}^{\infty} \ord \left( \snr^{i + j\mu} \right)} \nonumber \\
\etr & = & \tsty{ \frac{1}{2} \: \snr^{1-\mu} +
\sum_{i=1}^{\infty} \sum_{j=1}^{\infty} \ord \left( \snr^{i +
j\mu} \right)} \label{asympcase1}
\end{eqnarray}
which implies that $\etas \rightarrow \frac{1}{2}$ and $\mseq
\rightarrow 1$ (since $\etr \rightarrow 0$).

Case 2: When $\mu = 1$
\begin{eqnarray}
\eta_{1} & = & 1 + \tsty{ \ord \left( \snr \right) } \nonumber \\
\eta_{2} & = & \tsty{ \sqrt {2 + \ord \left( \snr \right) + \ord
\left( \snr^{2} \right)} - 1 }. \label{case2}
\end{eqnarray}
The above relationships imply that $\etas \rightarrow 0.414 \hsp
\text{and} \hsp \mseq \rightarrow 0.707$.

Case 3: For $\mu > 1$, $\eta_{1}$ is the same as in
(\ref{eta1case1}) but the asymptotic expansion for $\eta_{2}$ is
\begin{equation}
\eta_{2} = \snr^{\frac{1-\mu}{2}} - 1 + \littleo (1).
\label{eta2case3}
\end{equation}
It is easy to see in this case that $\etas \rightarrow 0$.
Similarly it follows that $\etr \rightarrow \infty$ and so $\mseq
\rightarrow 0$. Furthermore, the rates of convergence in this case
can be obtained using (\ref{eta1case1}) and (\ref{eta2case3}) and
is as illustrated in (\ref{rates}).
\endproof

\bibliographystyle{IEEEbib}
\bibliography{bib_jstsp2}

\end{document}